\documentclass[aps, prb, superscriptaddress, reprint, nobibnotes, floatfix]{revtex4-2}
\usepackage{amsmath}
\usepackage{color}
\usepackage{comment}
\usepackage{bm,graphicx,hyperref}
\hypersetup{%
  breaklinks = {true},
  citecolor = {blue},
  colorlinks = {true},
  linkcolor = {red},}

\begin{document}

\title{Microscopic theory of thermalization in 1D with nonlinear bath coupling}
\author{A. Rodin}
\thanks{Corresponding author}
\affiliation{Yale-NUS College, 16 College Avenue West, 138527, Singapore}
\affiliation{Centre for Advanced 2D Materials, National University of Singapore, 117546, Singapore}
\affiliation{Materials Science and Engineering, National University of Singapore, 117575, Singapore}

\author{B. A. Olsen}
\thanks{Corresponding author}
\affiliation{Yale-NUS College, 16 College Avenue West, 138527, Singapore}

\author{M. Choi}
\affiliation{Yale-NUS College, 16 College Avenue West, 138527, Singapore}

\author{A. Tan}
\affiliation{Yale-NUS College, 16 College Avenue West, 138527, Singapore}

\begin{abstract}

Using a non-perturbative classical model, we numerically investigate the dynamics of mobile particles interacting with an infinite chain of harmonic oscillators, an abstraction of ionic conduction through solid-state materials.
We show that coupling between the mobile particles and a single mass of the chain is sufficient to induce dissipation of the mobile particles' energy over a wide range of system parameters.
When we introduce thermal fluctuations in the position of the chain mass, the mobile particles exhibit thermalization, eventually reaching the same temperature scale as the chain.
This model demonstrates how a minimal set of ingredients can exhibit a link between microscopic motion and macroscopic observables, with computationally efficient simulations.
Finally, we suggest some experimental platforms that could realize such a model.

\end{abstract}  

\maketitle

\section{Introduction}
\label{sec:Introduction}

The relationship between microscopic particle motion and macroscopic system parameters lies at the heart of statistical mechanics, and has many implications both in basic science and technology.
The fluctuation-dissipation theorem (FDT), formally established by Callen and Welton~\citep{Callen1951}, demonstrated the link between viscous drag and thermal fluctuations experienced by a Brownian particle, first exposed by Einstein~\citep{Einstein1905} and Smoluchowski~\citep{Smoluchowski1906}.
In short, collisions between a Brownian particle and the medium hosting it drain the particle's energy, giving rise to drag and dissipation while also imparting energy to the particle due to the thermal motion of the medium.

In Brown's original experiments~\citep{Brown1828, Pearle2010}, the observed particles were pollen organelles colliding with water molecules. 
Because each Brownian particle is much heavier than the water molecules, the time scale for its motion is much longer. 
Therefore, the force exhibited on the particle by the water can be treated as uncorrelated white noise, an approximation which is expected to break down when the masses of the medium's constituent particles and the Brownian particle are similar. 
An example of such a situation is ionic motion through solid materials, where the mobile ions have similar masses to the ions constituting the lattice. 
Furthermore, the motion of the lattice ions is correlated due to long-range order.

Ionic transport in solids is garnering growing attention due to interest in developing solid-state batteries~\cite{Bachman2016, Manthiram2017, Famprikis2019}.
One of this technology's integral components is the solid-state electrolyte: an electronically insulating material that can conduct ions and serves as a separator between the anode and the cathode.
One recent approach proposed a microscopic theory to describe the dissipative motion of ions through crystalline solids~\citep{Rodin2021a}.
In the limit of high temperature and long time, the effect of correlations is severely diminished, and FDT can relate random thermal forces to drag in the material.

The problem of a small mobile particle coupled to a dissipative thermal bath has been of interest to the physics community for a long time~\citep{Feynman1963, Caldeira1981}.
In recent years, there have been significant advances in understanding the dynamics of impurities immersed in bosonic~\citep{Caldeira1995, Schecter2012, Peotta2013, Dehkharghani2015, Petkovic2016} and fermionic~\citep{Caldeira1995, CastroNeto1996, Pasek2019} systems.
In recent work, linearized approaches have been used to show the emergence of Brownian motion in $D$-dimensional Bose-Einstein Condensates~\citep{Lampo2017}, as well as the microscopic origins of friction in one-dimensional quantum liquids~\cite{Petkovic2020}.

In this work, we build on the formalism from Ref.~\citep{Rodin2021a} to approach ion transport from a classical perspective. 
We construct a minimal experimentally-realizable model to show how particles trapped in a harmonic potential and coupled to an `ion framework' composed of a one-dimensional chain of harmonic oscillators can exhibit both dissipation and thermalization. 
We demonstrate that interactions between the mobile particles and a single mass of the chain is sufficient to induce fluctuation-dissipation behavior. 
One advantage of our classical approach is that it does not rely on the assumption that the displacement of the chain masses is small.
In contrast, most quantum-mechanical formulations assume the coupling between the ion framework and the mobile particles is linear in the displacement. 

In Sec.~\ref{sec:General_Model}, we derive the integro-differential equations of motion for a collection of mobile particles travelling through a potential landscape generated by a vibrating lattice of arbitrary dimensionality. 
In Sec.~\ref{sec:1D_Chain}, we simplify those equations for a one-dimensional system with a specific lattice geometry, and describe the computational procedure for numerically simulating the particle trajectories. 
To disentangle the effects of dissipation and fluctuation, in Sec.~\ref{sec:Dissipation} we explore the behavior of systems at $T=0$.
We test the scaling of dissipation with system parameters, and discuss the role of the `memory' term arising from integrating out the chain degrees of freedom.
In Sec.~\ref{sec:Thermalization}, we study the role of fluctuations on particle trajectories by varying the temperature of the lattice. 
Finally, we propose some potential platforms for experimental validation of this model using cold atoms or ions in Sec.~\ref{sec:Implementations}. 

\section{General Model}
\label{sec:General_Model}

We begin by considering a general Lagrangian (in spatial dimension $D$) describing the motion of mobile particles through a framework of masses with vibrational modes,

\begin{equation}
    L = T_M(\dot{\mathbf{R}}) - V_M(\mathbf{R})
    + T_F(\dot{\mathbf{r}})  - V_F(\mathbf{r})
    - U(\mathbf{r},\mathbf{R}, t) \,.
    \label{eqn:Lagrangian}
\end{equation}
Here we combine the displacements of the framework masses from their equilibrium positions into a single vector $\mathbf{r} = \bigoplus_{j=1} \mathbf{r}_j$, and combine the mobile particle positions as $\mathbf{R}$. $T_M(\dot{\mathbf{R}})$ and $T_F(\dot{\mathbf{r}})$ are the kinetic energies of the mobile particles and framework masses, respectively, while $V_M(\mathbf{R})$ and $V_F(\mathbf{r})$ are the corresponding time-independent potential energies. Finally, $U(\mathbf{r}, \mathbf{R}, t)$ is a general potential energy that describes all remaining interactions and perturbations.

Assuming the homogeneous motion of the framework masses is harmonic, we write

\begin{align}
    T_F(\mathbf{r}) - V_F(\mathbf{r}) \rightarrow 
     \frac{1}{2}\dot{\mathbf{r}}^T\tensor{m}\dot{\mathbf{r}}
    -
    \frac{1}{2}\mathbf{r}^T \tensor{V} \mathbf{r}\,.
    \label{eqn:Harmonic_Lagrangian}
\end{align}
Here, $\tensor{m} = \bigoplus_{j=1} m_j\tensor{1}_{D}$ is a block-diagonal matrix where $m_j$ is the mass of the $j$th framework mass, and $\tensor{V}$ is the harmonic coupling matrix.

The homogeneous equation of motion $\tensor{m}\ddot{\mathbf{r}} =  -\tensor{V}\mathbf{r}$ obtained from Eq.~\eqref{eqn:Harmonic_Lagrangian} can be transformed into a symmetric eigenvalue problem by first defining $\tilde{\mathbf{r}} = \tensor{m}^{\frac{1}{2}}\mathbf{r}$ so that

\begin{equation}
    \ddot{\tilde{\mathbf{r}} }
    = 
    -\Omega_j^2\tilde{\mathbf{r}}
    = 
    -
    \tensor{m}^{-\frac{1}{2}}\tensor{V}\tensor{m}^{-\frac{1}{2}}\tilde{\mathbf{r}} 
    \equiv
    -
   \tilde{\tensor{V}}\tilde{\mathbf{r}} 
    \,,
    \label{eqn:Symmetric_EOM}
\end{equation}
with normalized eigenvectors $\boldsymbol{\varepsilon}_j$ and corresponding eigenvalues $\Omega_j^2$. Hence, we can write $\tilde{\mathbf{r}}(t) = \tensor{\varepsilon}\boldsymbol{\zeta}(t)$ and, consequently, $\mathbf{r}(t) =\tensor{m}^{-\frac{1}{2}} \tensor{\varepsilon}\boldsymbol{\zeta}(t)$, where $\boldsymbol{\zeta}(t)$ is a column vector of normal coordinates giving the amplitude of each mode, and $\tensor{\varepsilon} = [\boldsymbol{\varepsilon}_1, \boldsymbol{\varepsilon}_2, \dots]$ is a row of column vectors $\boldsymbol{\varepsilon}_j$.

Writing down the equations of motion for the framework masses using all the terms in Eq.~\eqref{eqn:Lagrangian} yields

\begin{align}
    \tensor{m}\ddot{\mathbf{r}} &= -\tensor{V}\mathbf{r} - \nabla_\mathbf{r} U(\mathbf{r},\mathbf{R},t)\,,
    \nonumber
    \\
    \rightarrow
    \ddot{\boldsymbol{\zeta}} &= -\tensor{\Omega}^2\boldsymbol{\zeta} - \tensor{\varepsilon}^{-1}\tensor{m}^{-\frac{1}{2}}\nabla_\mathbf{r} U(\mathbf{r},\mathbf{R},t) \,,
    \label{eqn:Inhomogeneous_EOM}
\end{align}
where $\tensor{\Omega}^2 = \tensor{\varepsilon}^{-1}\tilde{\tensor{V}}\tensor{\varepsilon}$ is a diagonal matrix of the squared eigenfrequencies. For a single normal coordinate, the equation of motion takes the form $\ddot{\zeta}_j = - \Omega_j^2 \zeta_j- f_j$, which can be solved using the Green's function formalism. Recalling that the Green's function for a harmonic oscillator is given by

\begin{equation}
     G_j(t,t')= \frac{\sin\left[\Omega_j(t - t')\right]}{\Omega_j}\Theta(t - t')\,,
     \label{eqn:Harmonic_Greens_fn}
\end{equation}
we have

\begin{align}
    \zeta_j(t)
   = \zeta_j^H(t) &- \int^t dt'\frac{\sin\left[\Omega_j(t - t')\right]}{\Omega_j}
   \nonumber
   \\
   &\times
   \left[\tensor{\varepsilon}^{-1}\tensor{m}^{-\frac{1}{2}}\nabla_\mathbf{r}U(\mathbf{r},\mathbf{R},t')\right]_j 
   \nonumber
   \\
   = \zeta_j^H(t) &- \int^t dt'\frac{\sin\left[\Omega_j(t - t')\right]}{\Omega_j}
   \nonumber
   \\
   &\times
   \boldsymbol{\varepsilon}^T_j\tensor{m}^{-\frac{1}{2}}\nabla_\mathbf{r}U(\mathbf{r},\mathbf{R},t')
   \,,
   \label{eqn:zeta_solution}
\end{align}
where $\zeta_j^H(t)$ is the homogeneous solution and the subscript $j$ at the braces indicates that we pick out the $j$th element of the column vector. The second equality follows from the fact that $\tensor{\varepsilon}$ is an orthogonal matrix so that $\tensor{\varepsilon}^{-1} = \tensor{\varepsilon}^T$. Using $\mathbf{r}(t) =\tensor{m}^{-\frac{1}{2}} \sum_j \boldsymbol{\varepsilon}_j\zeta_j(t) $, we obtain

\begin{align}
    \mathbf{r}(t)
   &=  \tensor{m}^{-\frac{1}{2}}\sum_j\boldsymbol{\varepsilon}_j\zeta_j^H(t) 
   \nonumber
   \\
   &-
   \int^t dt'
   \tensor{m}^{-\frac{1}{2}}
   \tensor{G}(t - t')
   \tensor{m}^{-\frac{1}{2}}\nabla_\mathbf{r}U(\mathbf{r},\mathbf{R},t')
   \,,
   \label{eqn:r_solution}
   \\
   \tensor{G}(\tau) & = 
   \sum_j \boldsymbol{\varepsilon}_j\otimes \boldsymbol{\varepsilon}_j\frac{\sin\left(\Omega_j\tau\right)}{\Omega_j}\,.
   \label{eqn:G_mat}
\end{align}

The homogeneous solution $\zeta_j^H(t)$ is determined by the thermodynamic properties of the framework.
To relate the amplitudes of the modes to the framework temperature, 
we consider the Lagrangian for a single normal mode $\zeta_j$, given by $L_j = \dot{\zeta}_j^2/2 - \Omega^2 \zeta_j^2/2$. 
The solution to the resulting equation of motion is $\zeta_j(t) = A_j\cos(\Omega_j t + \phi_j)$,  where $0\leq\phi_j<2\pi$ is a phase factor determined by boundary conditions.

To generate the $\mathbf{r}(t)$ originating from the thermal motion, we obtain a set of $\phi_j$ and $A_j$ that correctly reflect the system's thermodynamics.
The phases $\phi_j$ are sampled from a uniform distribution $[0, 2\pi)$.
To generate the amplitudes $A_j$, we recall that the amplitude is related to the total energy of the oscillator mode. 
For ease of calculation, we treat the possible energies as discrete, following the solution of the quantum mechanical harmonic oscillator.
Then the amplitude becomes a function of the number of quanta $n$: $A_j(n)$.
The resulting expectation value of the square of the displacement is

\begin{align}
    \langle \zeta_{j}(t)\zeta_{j}(t)\rangle
    &=
    \oint \frac{d\phi_j}{2\pi}
    \frac{\sum_n A_j^2(n)\cos^2(\Omega_j t + \phi_j) e^{-n\Omega_j / \Omega_T}}{\sum_ne^{-n\Omega_j / \Omega_T}}
    \nonumber
    \\
    &=
   \frac{1}{2}
    \frac{\sum_n A_j^2(n) e^{-n\Omega_j / \Omega_T}}{\sum_ne^{-n\Omega_j / \Omega_T}}
    \,,\label{eqn:zeta_squared}
\end{align}
where $n_B$ is the Bose-Einstein distribution and $\Omega_T = k_B T / \hbar$ is the thermal frequency.
Recalling the familiar result for a quantum harmonic oscillator $\langle \zeta_{j}(t)\zeta_{j}(t)\rangle = \frac{\hbar}{\Omega_j}
    \left[n_B(\Omega_j)
    +\frac{1}{2}
    \right]$, 
we find that  $A_j(n_j) = \sqrt{n_j + \frac{1}{2}}\sqrt{\frac{2\hbar}{\Omega_j}}$, where $n_j$ is an integer obtained from the probability distribution $e^{-n\Omega_j / \Omega_T}$.

The second term in Eq.~\eqref{eqn:r_solution} encapsulates all of the interactions in the system.
Since this term integrates a generalized force for all past times, we refer to it as the `memory' term in the trajectory, and $\tensor{G}$ as the `memory kernel'.
While formally the memory is integrated over all past times, we will explore the consequences of truncating the integration in Sec.~\ref{sec:Role_of_Memory}.

The advantage of the memory formalism becomes evident when writing down the the equations of motion for the mobile particles,

\begin{equation}
    \tensor{M}\ddot{\mathbf{R}} = -\nabla_\mathbf{R}\left[U\left(\mathbf{r},\mathbf{R},t'\right)+V_M\left(\mathbf{R}\right)\right]\,,
    \label{eqn:Mobile_EOM}
\end{equation}
where $\tensor{M} = \bigoplus_{j} M_j\tensor{1}_{D}$ is the equivalent of $\tensor{m}$ for the mobile particles. 
In order to obtain the trajectories of the mobile particles, we see that we can neglect the components of $\mathbf{r}$ which do not appear in $U(\mathbf{r},\mathbf{R},t)$. 
With this restriction, Eq.~\eqref{eqn:r_solution} allows us to solve only the relevant components of $\mathbf{r}$ and ignore other degrees of freedom.
This simplification becomes more dramatic if the mobile particles interact with a small fraction of the framework masses.
In this way, if the number of extraneous degrees of freedom of the framework is large enough, it can act as a thermal bath able to exchange energy with the mobile particles via the interaction $U$.

\section{1D Chain}
\label{sec:1D_Chain}

\subsection{Problem Formulation}
\label{sec:Problem Formulation}

\begin{figure}
    \centering
    \includegraphics[width=\columnwidth]{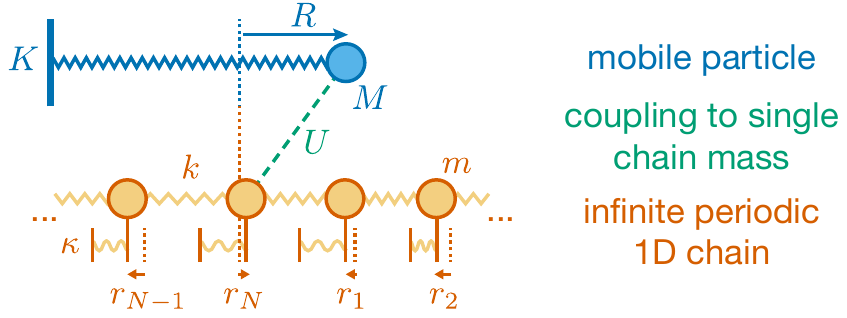}
    \caption{\textbf{Schematic of the system.} A mobile particle of mass $M$ undergoes 1D harmonic motion with spring constant $K$ and displacement from equilibrium $R$. It couples via potential $U$ to the $N$th mass from a periodic 1D chain of $N\rightarrow \infty$ masses, each of which undergoes harmonic motion with spring constant $\kappa$ and displacement $r_g$ and couples to its neighbor with spring constant $k$. The 1D chain has $N$ modes of finite bandwidth and acts as a bath, so the coupling leads to fluctuation and dissipation of the mobile mass's motion.}
    \label{fig:sketch}
\end{figure}
The simplest system with sufficient extraneous degrees of freedom to act as a heat bath is shown in Fig.~\ref{fig:sketch}.
In this setup, the framework is composed of a periodic chain of $N\rightarrow \infty$ identical masses $m$ connected by identical springs with force constant $k$ and restricted to one-dimensional motion.
Each chain mass is also confined by an external harmonic potential with force constant $\kappa$ to suppress zero-frequency modes which can cause instabilities in low-dimensional systems. 
The vibrational eigenmodes have frequencies
\begin{align}
    \Omega_j &= \sqrt{\frac{\kappa}{m}+4\frac{k}{m}\sin^2\left(\frac{\pi j}{ N}\right)}
    \nonumber
    \\
    &= \sqrt{\Omega_\mathrm{min}^2 \cos^2\left(\frac{\pi j}{ N}\right)+\Omega_\mathrm{max}^2 \sin^2\left(\frac{\pi j}{ N}\right)}
    \label{eqn:Omega}
\end{align}
with corresponding normalized eigenvectors $\varepsilon_{g,j}^{\cos} = \cos(q_j g) \sqrt{2/N}$ and $\varepsilon_{g,j}^{\sin} = \sin(q_j g) \sqrt{2/N}$ for $q_j = 2\pi j / N$ with $1\leq j\leq N/2$, where $1\leq g\leq N$ is the index of the chain particle. Here,
$\Omega_\mathrm{max} = \sqrt{4k/m+\kappa/m}$ and $\Omega_\mathrm{min} = \sqrt{\kappa / m}$ are the maximum and minimum frequencies of the eigenmodes, which form an acoustic phonon band.

The mobile particles, each of mass $M$, move in 1D parallel to the chain and experience a harmonic potential with force constant $K$. The minimum of this potential coincides with the minimum of the harmonic well containing the $N$th chain particle, as shown in Fig.~\ref{fig:sketch}. 
The mobile particles do not interact with each other, so $V_M  = \frac{1}{2}K\sum_jR_j^2$ in Eq.~\eqref{eqn:Mobile_EOM}. In addition, the interactions are restricted to pairwise couplings between the chain mass and the mobile particles and do not contain an explicit time dependence, allowing us to write $U(\mathbf{r}, \mathbf{R}, t)$ as $U(r_N, \mathbf{R}) =  \sum_j U(r_N, R_j)$.

Because only the $N$th chain particle interacts with the mobile particles, it is the only mass whose position is relevant to the system dynamics. Consequently, $\mathbf{r}$ in Eq.~\eqref{eqn:r_solution} contains a single entry $r_N$. Moreover, we retain only the $N$th element of the eigenvectors $\varepsilon_{N,j}^{\cos} = \sqrt{2/N}$, $\varepsilon_{N,j}^{\sin} = 0$, resulting in

\begin{align}
    r_N(t)
   &=  \sqrt{\frac{2}{N}}\frac{1}{\sqrt{m}}\sum_j\sqrt{n_j + \frac{1}{2}}\sqrt{\frac{2\hbar}{\Omega_j}}\cos(\Omega_j t + \phi_j)
   \nonumber
   \\
   &-
   \frac{1}{m}
   \int^t dt'
   G(t - t')
    \frac{d U\left[r_N(t'),\mathbf{R}(t')\right]}{dr_N}
   \,,
   \label{eqn:r_N}
   \\
   G(t) & = 
   \frac{2}{N}
   \sum_{j=1}^{N/2} \frac{\sin\left(\Omega_jt\right)}{\Omega_j}
   \nonumber
   \\
   &=\frac{2}{\pi}\int_{\Omega_\mathrm{min}}^{\Omega_\mathrm{max}}dz \frac{\sin\left(t z\right)}{\sqrt{z^2 - \Omega^2_\mathrm{min}}\sqrt{\Omega^2_\mathrm{max} - z^2 }}\,,
   \label{eqn:G_1D}
   \\
   \ddot{R}_{j}(t) &=\frac{1}{M} \left\{-\frac{d}{dR_j}U\left[r_N(t),R_{j}(t)\right] - K R_{j}(t)\right\}\,.
   \label{eqn:R_j}
\end{align}

Since we are interested primarily in the motion of the mobile particles, we define several characteristic scales: $\Omega_M = \sqrt{K/M}$ is the homogeneous oscillation frequency of the mobile particles, with period $t_M = 2\pi/\Omega_M$ and energy $E_M = \hbar \Omega_M = Kl^2_M$, where $l_M = \sqrt{\hbar/M\Omega_M}$ is the quantum oscillator length. Rewriting Eqs.~\eqref{eqn:r_N}-\eqref{eqn:R_j} in terms of these characteristic quantities and expressing $m$ as a multiple of $M$ yields

\begin{align}
    \rho(\tau)
   &=  \sqrt{\frac{2}{N}}\sum_j\sqrt{n_j + \frac{1}{2}}\sqrt{\frac{2}{\mu\omega_j}}\cos(2\pi\omega_j \tau + \phi_j)
   \nonumber
   \\
   &-
   \frac{2\pi}{\mu}
   \int^\tau d\tau'
   \Gamma(\tau - \tau')
    \sum_j\frac{d \Phi\left[\rho(\tau'),\sigma_j(\tau')\right]}{d\rho}
   \,,
   \label{eqn:r_N_unitless}
   \\
   \Gamma(\tau) &= \frac{2}{\pi}\int_{\omega_\mathrm{min}}^{\omega_\mathrm{max}}dx \frac{\sin\left(2\pi\tau x\right)}{\sqrt{x^2 - \omega^2_\mathrm{min}}\sqrt{\omega^2_\mathrm{max} - x^2}}
   \,,
   \label{eqn:G_1D_unitless}
   \\
   \ddot{\sigma}_{j}(\tau) &=(2\pi)^2 \left\{-\frac{d}{d\sigma_j}\Phi\left[\rho(\tau),\sigma_{j}(\tau)\right] - \sigma_{j}(\tau)\right\}\,.
   \label{eqn:R_j_unitless}
\end{align}
Note that we dropped the subscript $N$ because it is redundant as we are keeping track of a single chain mass.
To help keep track of the correspondence between regular and dimensionless parameters of the model, we collect them in Table~\ref{tab:unitless}.

Despite the simplifications due to the 1D geometry, Eqs.~\eqref{eqn:r_N_unitless}-\eqref{eqn:R_j_unitless} are still not tractable analytically and require a numerical approach to compute the trajectories.

\begin{table}[tb]
    \centering
    \begin{tabular}{c||c|rl}
    Parameter&     & \multicolumn{2}{c}{Dimensionless}\\
    \hline \hline
     chain mass position & $r$ & $\rho$ & $=r/l_M$ \\
     mobile mass position & $R$ & $\sigma$ & $=R/l_M$ \\
     evolution time & $t$ & $\tau$ & $=t/t_M$ \\
     chain particle mass & $m$ & $\mu$ & $=m/M$ \\
     frequency & $\Omega$ & $\omega$ & $=\Omega/\Omega_M$ \\
     interaction strength & $U$ & $\Phi$ & $=U/E_M$ \\
     memory kernel & $G$ & $\Gamma$ & $=G \cdot\Omega_M$ \\
     interaction length & $s$ & $\lambda$ &$=s/l_M$ \\
     mobile particle energy & $E$ & $\mathcal{E}$ & $=E/E_M $\\
     thermal frequency & $\Omega_T$ & $\omega_T$ & $= k_BT/E_M$
    \end{tabular}
    \caption{\textbf{Parameters used in the model.} In the simulations, we formulate the dynamics using dimensionless quantities, given in the right column. This formulation leads to values for most quantities on the order of unity, which helps avoid numerical issues due to machine precision. Here, $\Omega_M = \sqrt{K/M}$, $t_M = 2\pi/\Omega_M$, $E_M = \hbar \Omega_M = Kl^2_M$, and $l_M = \sqrt{\hbar/M\Omega_M}$.}
    \label{tab:unitless}
\end{table}

\subsection{Computational Procedure}
\label{sec:Computational_Procedure}

Numerically integrating Eqs.~\eqref{eqn:r_N_unitless} and \eqref{eqn:R_j_unitless}, while conceptually straightforward, can be computationally demanding.
We performed our computations using the {\scshape julia} programming language~\citep{Bezanson2017}, and our code is available at \footnote{https://github.com/rodin-physics/1d-parabolic-trap-thermalization}. {\scshape julia} is well-suited for scientific computing due to a number of native optimizations.
All our plots are visualized using {\scshape Makie.jl} ~\cite{DanischKrumbiegel2021} and employ a scheme suitable for color-blind readers, developed in \citep{wong_points_2011}.

Here we describe the computational techniques that complement the scripts at \footnote{https://github.com/rodin-physics/1d-parabolic-trap-thermalization} to reproduce all the calculations presented in the following sections.
We do not provide the output files because of their size, but we do include the scripts used to generate them.

By rewriting Eqs.~\eqref{eqn:r_N_unitless} and \eqref{eqn:R_j_unitless} using discrete time steps $\delta$ so that $\tau=\delta\alpha$ for integer $\alpha$'s, we have 

\begin{align}
    \rho_\alpha
   &= \sqrt{\frac{2}{N}}\sum_j\sqrt{n_j + \frac{1}{2}}\sqrt{\frac{2}{\mu\omega_j}}\cos(2\pi\omega_j \delta \alpha + \phi_j)
   \nonumber
   \\
   &-
   \frac{2\pi\delta}{\mu}
   \sum_{\beta}
   \Gamma\left[\delta(\alpha - \beta)\right]
    \sum_j\frac{d \Phi\left(\rho_{\beta},\sigma_{j,\beta}\right)}{d\rho}
   \,,
   \nonumber
   \\
   \sigma_{j,\alpha} &= (2\pi \delta)^2 \left[-\frac{d}{d\sigma_j}\Phi\left(\rho_{\alpha-1},\sigma_{j,\alpha -1}\right) - \sigma_{j, \alpha - 1}\right]
   \nonumber
   \\
   &+2\sigma_{j, \alpha - 1}-\sigma_{j, \alpha - 2}\,.
   \label{eqn:EOM_discrete}
\end{align}
These difference equations can be solved using iteration, as is common for initial value problems. In our case, we initialize $\sigma_{j,0}$ and $\sigma_{j,1}$ to the same value to have the mobile particles start from rest.

To guarantee the smoothness of the solution, it is important to consider two factors. 
First, the time step $\delta$ has to be much smaller than the period of the fastest chain mode. 
Second, $\delta$ has to be sufficiently small so that the force experienced by the rapidly moving mobile particles is smooth in the vicinity of the chain particle with which they interact. We will illustrate the fulfillment of these conditions in the following section.

Because $\rho_\alpha$ and $\sigma_{j,\alpha}$ in Eq.~\eqref{eqn:EOM_discrete} depend on the earlier positions, the solution of the difference equation is not parallelizable, which slows down the calculation. 
Fortunately, we can alleviate some of the computational load by precomputing the memory kernel $\Gamma$. 
We first define the time period for the simulation $\tau \in [0,\tau_f]$ and partition this period into steps of size $\delta$, where $\delta$ is chosen to satisfy the requirements described above. 
Next, we calculate an array $[\Gamma(\delta), \Gamma(2\delta),\dots ,\Gamma(\tau_f)]$ by integrating Eq.~(\ref{eqn:G_1D_unitless}) using the Gaussian quadrature method. Because the entries of the array are independent of one another, they can be computed in parallel. Precomputing the memory kernel eliminates the slow integration step from the sequential solution of Eq.~\eqref{eqn:EOM_discrete}, leaving only algebraic calculations.

In the course of the simulation, we save the force terms $\sum_j\frac{d}{d\rho}\Phi\left(\rho_\alpha, \sigma_{j,\alpha}\right)$ and $\frac{d}{d\sigma_j}\Phi\left(\rho_{\alpha-1},\sigma_{j, \alpha-1}\right)$ for each time step. 
We can then calculate the memory term by multiplying the `past' forces by the appropriate entries of the precomputed $\Gamma$ array and performing a summation. 
Here $\Gamma(0)$ is multiplied by the current force, $\Gamma(\delta)$ by the force from the previous step, $\Gamma(2\delta)$ by the force from two steps ago, and so on.
That is, $\Gamma$'s with larger time arguments get multiplied by `older' forces.

Although the exact form of the interaction between mobile particles and the chain mass will lead to some quantitative differences in the particle trajectory, we expect the qualitative behavior to be independent of the details of the potential. 
Therefore, for the sake of simplicity, we choose 
$\Phi=\Phi_0 \exp\left[-(\rho_\alpha-\sigma_{j,\alpha})^2/2\lambda^2\right]$, where $F=\hbar \Omega_M \Phi_0$ is the interaction amplitude, and $s=\lambda l_M$ is the characteristic length scale of the interaction.

\begin{figure*}
    \centering
    \includegraphics[width=\textwidth]{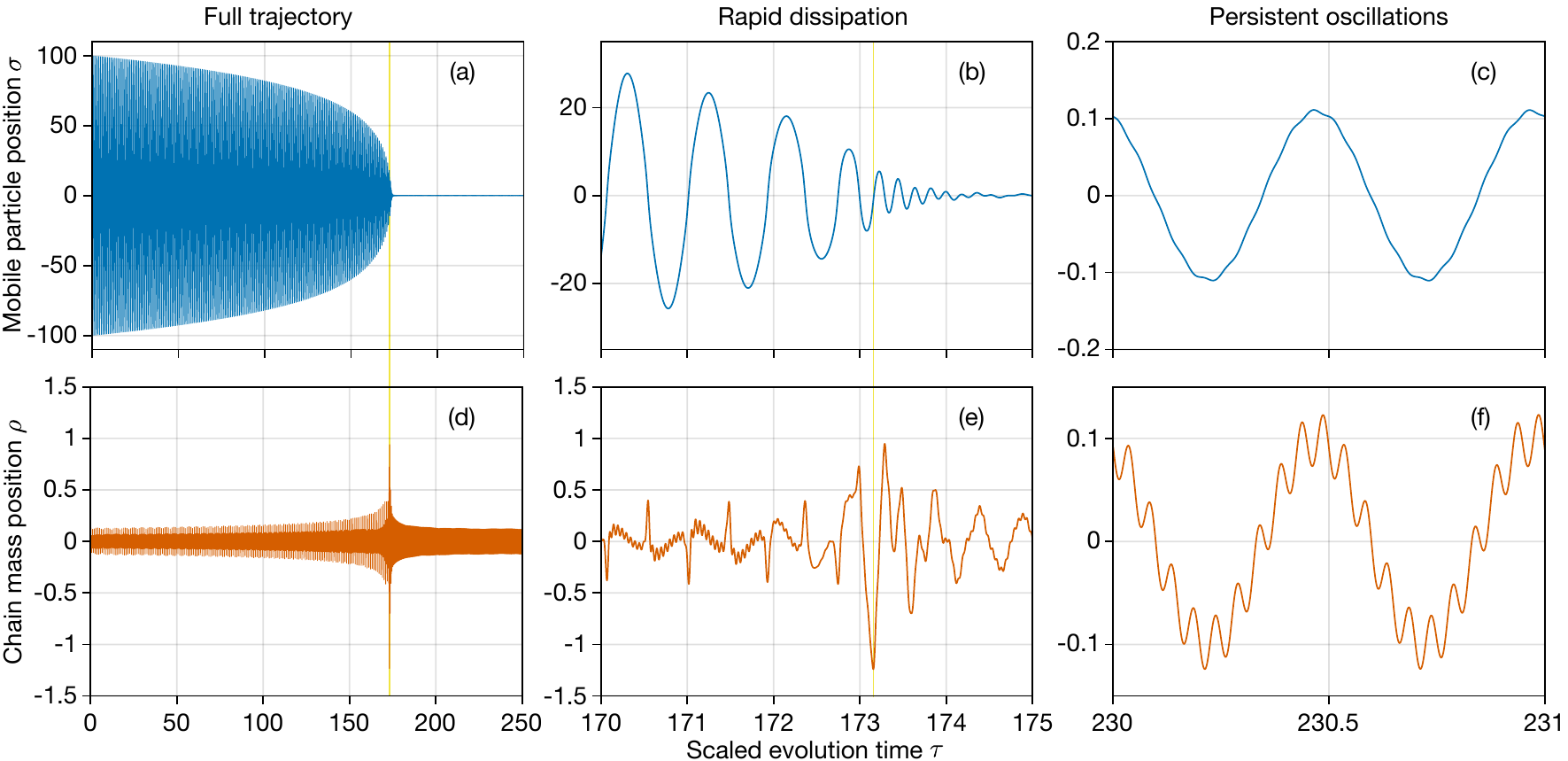}
    \caption{\textbf{General example with a single mobile particle.} Trajectories $\sigma(\tau), \rho(\tau)$ for the mobile particle and the interacting chain particle, respectively. Here we chose $\omega_\text{min}=2$, $\omega_\text{max}=20$, $\mu = 2$, $\lambda=4$, $\Phi_0 = -500$ with $\sigma(0) = 100$ and $\rho(0) = 0$. The mobile particle undergoes many oscillations and its energy slowly dissipates until about $\tau = 172$. Dissipation then quickly reduces its energy to nearly zero, and it falls into the potential well of the interacting chain particle, which reaches its largest displacement at around $\tau=173$, indicated by the vertical yellow line. For late times $\tau>175$, both particles undergo small, persistent, non-dissipative oscillations at two frequencies just outside the phonon band of the chain.}
    \label{fig:General_Example}
\end{figure*}

The final ingredient of the simulation is the homogeneous, or thermal, motion of the chain particle. 
To compute the trajectory for $10^6$ chain masses, we generate $5\times10^5$ equally-spaced values $q_j$ between $0$ and $\pi$, along with the corresponding frequencies $\omega_j$ given by Eq.~\eqref{eqn:Omega}. 
Next, we generate $5\times10^5$ random phases $0\leq \phi_j \leq 2\pi$, as well as random integers $n_j$ from the probability distribution $e^{-n \omega_j/\omega_T}$.
These values of $n_j$  and $\phi_j$ are then used to construct the trajectory $\rho^H(\tau)=\sum_j\sqrt{n_j + \frac{1}{2}}\sqrt{\frac{4}{N\mu\omega_j}}\cos(2\pi\omega_j \tau + \phi_j)$, as shown in Eq.\eqref{eqn:r_N_unitless}.


\section{Dissipation}
\label{sec:Dissipation}

\subsection{General Picture}
\label{sec:General_Picture}

To develop a better feel for the system behavior, we will begin by studying trajectories with vanishing thermal motion, $\rho^H=0$.
In this case, we expect the mobile particles to dissipate energy over time due to their interaction with the chain. 

As a first example, we choose a configuration with a single mobile particle and set $\omega_\text{min}=2$, $\omega_\text{max}=20$, $\mu=2$, $\lambda = 4$, and $\Phi_0=-500$ (attractive interaction). 
The trajectory for these parameters is shown in Fig.~\ref{fig:General_Example}.
We chose values for the system parameters partly for convenience (with most parameters on the order of unity), and partly to remain well in the classical regime (by keeping lengths and energies large compared to characteristic scales set by the harmonic oscillator).
We set the time step $\delta$ of the simulation to be substantially smaller than the period of the fastest chain mode $\tau_\mathrm{max} = \left(2\pi / \Omega_\mathrm{max}\right) / \left(2\pi / \Omega_M\right) = 1/\omega_\mathrm{max}$. 
When written in terms of dimensionless quantities, periods and frequencies satisfy $\omega\tau=1$ (while $\Omega t  = 2 \pi$).
For our calculations, $\delta = 1 /60 \omega_\mathrm{max}$.




In Fig.~\ref{fig:General_Example}, we can identify two qualitatively distinct regimes. 
For $\tau < 172$, the mobile particle essentially undergoes simple harmonic motion with gradually decreasing amplitude.
Because the oscillation amplitude is much larger than the range of the interaction, $\lambda$, the mobile particle spends very little time interacting with the chain, which explains why the oscillations are almost sinusoidal.
The amplitude of the chain mass motion grows slowly.
Intuitively, when the mobile particle passes the chain mass, the latter is displaced due to the interaction term. As the amplitude of the mobile particle's motion decreases, the speed with which it passes the chain mass become smaller. Consequently, the time of the interaction grows, leading to an increase of the chain mass displacement.

For $\tau > 175$, the motion of the two particles appears to be a superposition of two modes with the slower one being in-phase for the two objects and the faster one being completely out-of-phase. 
Moreover, the amplitude of the oscillations appears to persist over many cycles, suggesting a lack of energy dissipation. 
To confirm that this non-dissipative motion is not a numerical artifact, we explore the long-term system behavior when $|\rho - \sigma|\ll \lambda$.
In this case, the Gaussian interaction term can be expanded to yield 

\begin{align}
    \rho (\tau)
    &=
    \frac{2\pi\Phi_0}{\mu\lambda^2}\int^\infty_{-\infty} d\tau'
    \Theta(\tau - \tau')
    \nonumber
    \\
    &\times 
    \frac{2}{N}
   \sum_{j=1}^{N/2} \frac{\sin\left[2\pi\omega_j(\tau - \tau')\right]}{\omega_j}\left[\rho(\tau') - \sigma(\tau')\right]\,,
    \nonumber
    \\
   \ddot{\sigma}(\tau) &=(2\pi)^2 \left\{\frac{\Phi_0}{\lambda^2}\left[\sigma(\tau) - \rho(\tau)\right] - \sigma(\tau)\right\}\,,
   \label{eqn:Small_Amp_Osc}
\end{align} 
where we extended the lower time limit to $-\infty$ to focus on the long-term behavior.
Taking the Fourier transform of this system of equations using the definition $g_\omega =\mathcal{F}\left[g(\tau)\right] = \left(2\pi\right)^{-1/2}\int_{-\infty}^\infty d\tau e^{i2\pi\omega \tau} g(\tau)$ gives

\begin{align}
    \rho_\omega
    &=
    \frac{2\pi\Phi_0}{\mu\lambda^2}
    \frac{2}{N} \sum_{j=1}^{N/2}
    \Big[\frac{i\pi}{2\omega_j}\left[\delta(\omega -  \omega_j)-\delta(\omega +  \omega_j)\right]
    \nonumber
    \\
    &-\frac{2\pi}{4\pi^2\omega^2-4\pi^2\omega_j^2}\Big]\left(\rho_\omega - \sigma_\omega\right)\,,
    \nonumber
    \\
   -(2\pi\omega)^2\sigma_\omega &=(2\pi)^2 \left[\frac{\Phi_0}{\lambda^2}\left(\sigma_\omega - \rho_\omega\right) - \sigma_\omega\right]\,.
   \label{eqn:Fourier}
\end{align} 
We included the factor of $2\pi$ inside the exponential to agree with the argument form of the memory kernel. 
From the Fourier transform, we see that only if $\omega \notin \left[\omega_\mathrm{min}, \omega_\mathrm{max}\right]$, the delta functions both vanish and the system of equations may admit persistent real-$\omega$ solutions. 
Therefore, we drop the delta functions to get $\rho_\omega=\frac{\Phi_0}{\mu\lambda^2}\left(\sigma_\omega - \rho_\omega\right)f_\omega$ with
\begin{equation}
    f_\omega =  \frac{2}{N} \sum_{j=1}^{N/2}
    \frac{1}{\omega^2-\omega_j^2} 
    =\frac{1}{\omega^2 - \omega_\mathrm{max}^2}\sqrt{\frac{\omega^2 - \omega_\mathrm{max}^2}{\omega^2 - \omega_\mathrm{min}^2}}\,.
    \label{eqn:f_omega}
\end{equation}
Eliminating $\rho_\omega$ and $\sigma_\omega$ from the system of equations yields
\begin{equation}
    (1 - \omega^2)\left(1+\frac{f_\omega\Phi_0}{\mu \lambda^2}\right) = \frac{\Phi_0}{\lambda^2}\,.
    \label{eqn:small_osc_omega}
\end{equation}

Solving Eq.~\eqref{eqn:small_osc_omega} for the parameters used in the simulation reveals that there indeed are two persistent modes: one at $\omega\approx 0.999\omega_\mathrm{min}$ and another at $\omega\approx 1.001\omega_\mathrm{max}$. Because these two modes are outside the phonon band, they do not couple with the chain modes and, therefore, do not dissipate energy. 
This effect persists even when the $\rho^H(\tau)$ term is included, provided the random thermal noise is sufficiently weak.
Raising the temperature further will disrupt this periodic motion, as we will see below.

\subsection{Dissipation Scaling}
\label{sec:Dissipation_Scaling}

In the previous section, we discussed dissipation of a single mobile particle's energy for a particular choice of system parameters.
We now address how the choice of these parameters determines the dissipation rate.
To make analytical progress, we focus on the regime where the amplitude of the mobile particle's oscillations is substantially larger than the extent of the interaction potential.

To estimate the amount of energy that the mobile particle loses after a single encounter with the chain mass, we neglect the confining potential and assume that the particle travels at a constant speed from negative infinity so that $\sigma(\tau) = \dot{\sigma}_0\tau$.
This simplification is reasonable if the gain in the kinetic energy due to the harmonic trap in the vicinity of the chain mass is negligible. 
Next, we assume that the displacement of the chain mass is sufficiently small so that it can be dropped from the integral in Eq.~\eqref{eqn:r_N_unitless}, leading to

\begin{align}
    \rho (\tau)
    &= \frac{2\pi}{\mu}
    \int^\tau_{-\infty} d\tau'
    \Gamma\left(\tau - \tau'\right)
    \frac{d}{d\sigma}\Phi\left[\sigma(\tau')\right]\,,
    \label{eqn:r_reduced}
\end{align} 
where we used $d\Phi/d\rho = -d\Phi/d\sigma$.

To estimate the speed of the chain mass at $\tau = 0$ (as the mobile particle passes the origin), we differentiate Eq.~\eqref{eqn:r_reduced} with respect to $\tau$:

\begin{align}
    \dot{\rho} (0)
    &=
    \frac{2\pi}{\mu}\int^0_{-\infty} d\tau'
    \dot{\Gamma}\left( - \tau'\right)
    \frac{d\Phi}{d\sigma}\left(\dot{\sigma}_0 \tau'\right)
    \nonumber
    \\
    &=
    \frac{4\pi^2}{\mu}
    \int^0_{-\infty} d\tau'
    \frac{d\Phi}{d\sigma}\left(\dot{\sigma}_0 \tau'\right) = \frac{4\pi^2\Phi(0)}{\dot{\sigma}_0 \mu}\,.
    \label{eqn:r_dot}
\end{align} 
Here, we assume that the time during which the two particles interact is short compared to the periods of the chain modes, allowing us to replace $\dot{\Gamma}\left(-\tau'\right)\rightarrow \dot{\Gamma}\left(0\right)= 2\pi$.

\begin{figure*}
    \centering
    \includegraphics[width = \textwidth]{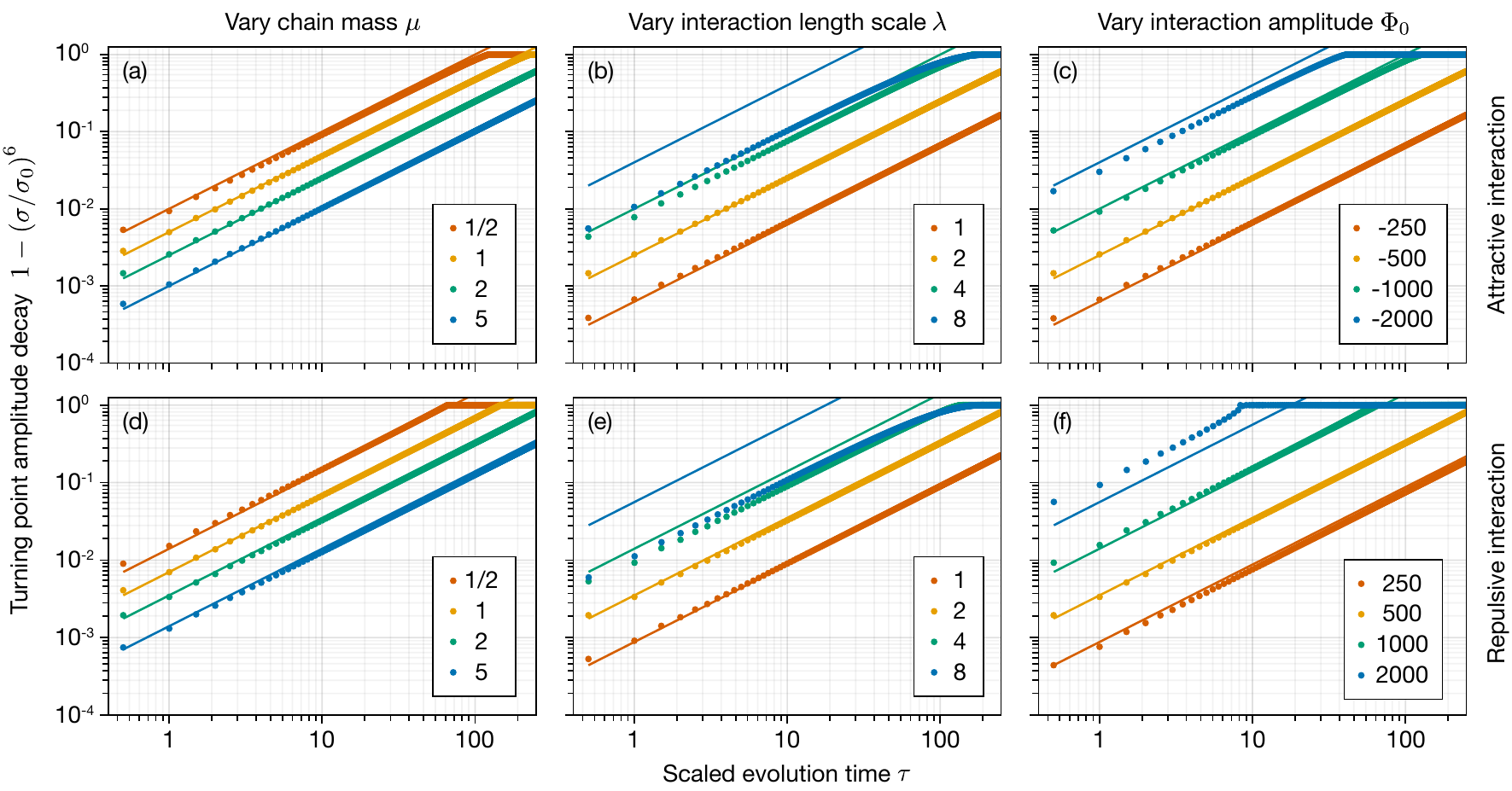}
    \caption{\textbf{Scaling of amplitude decay.} The amplitude of the mobile particle at its turning points as a function of time shows a quasi-power-law behavior:  $1-[\sigma(\tau)/\sigma_0]^6 = C\tau(\Phi_0\lambda)^2/\mu$. The dots show data from the numerical simulations and the lines are fits. All the lines in the top (bottom) row have the same fit parameter $C = 5\times 10^{-9}$ ($C = 7\times 10^{-9}$) and slopes of $+1$ on the log-log axes. The fixed parameters for all the panels are $\omega_\mathrm{min} = 2$ and $\omega_\mathrm{max} = 20$. (a) and (d) $\lambda  = 2$, $\Phi_0 = \mp500$; (b) and (e) $\mu = 2$, $\Phi_0 = \mp500$; (c) and (f) $\mu = 2$, $\lambda = 2$.}
    \label{fig:Power_Law}
\end{figure*}

The displacement of the chain mass is approximately $\dot{\rho}(0)\tau_\text{int} \sim \Phi(0) \lambda_\text{int} / (\dot{\sigma}_0^2\mu)$, where $\tau_\text{int} \sim \lambda_\text{int} / \dot{\sigma}_0$ is the length of time during which the two particles interact and $\lambda_\text{int}$ is the characteristic width of the interaction potential. 
Because the chain mass is confined by a potential well and springs connecting it to its neighbors, the potential energy associated with its displacement is approximately proportional to the displacement squared $\sim k_\mathrm{eff} \left[\dot{\rho}(0)\tau_\text{int}\right]^2$. 
Here $k_\mathrm{eff}$ is an effective spring constant determined by $k$, $\kappa$, and $\mu$. 
In our dimensionless formulation, we fix $\omega_\text{min} = \sqrt{(\kappa/\mu)/K}$ and $\omega_\text{max} = \sqrt{(4k/\mu)/K + (\kappa/\mu)/K}$, while allowing $\mu$ to change.
Therefore, to vary the parameter $\mu$, we must simultaneously vary $k$, $\kappa$, and hence $k_\text{eff}$, such that $k_\mathrm{eff}\propto \mu$.
Thus, the energy stored in the compressed spring is $\sim\mu \left[\dot{\rho}(0)\tau_\text{int}\right]^2$. 
This energy, originating from the moving particle, will be dissipated by the infinite chain and gives the energy loss of the mobile particle during a single pass: $\Delta \mathcal{E} \sim \mu[\Phi(0)\lambda_\text{int}]^2 / (\dot{\sigma}_0^4\mu^2) \sim [\Phi(0)\lambda_\text{int}]^2 / \mu\mathcal{E}^2$, where $\mathcal{E}=E/\hbar\Omega_M\sim \dot{\sigma}_0^2$. Because the frequency of the encounters between the two particles is virtually constant owing to the harmonic trap, the average energy loss rate is given by $\dot{\mathcal{E}}\sim-\omega_M \Delta \mathcal{E}= - [\Phi(0)\lambda_\text{int}]^2/\mu \mathcal{E}^2$, which yields 

\begin{equation}
    \mathcal{E}(\tau) = \left[\mathcal{E}_0^3-C\tau\frac{\Phi^2(0)\lambda_\text{int}^2}{\mu}\right]^\frac{1}{3}\,,
    \label{eqn:E_Decay}
\end{equation}
where $C$ is a numerical constant.

To illustrate the quasi-power-law behavior exhibited in Eq.~\eqref{eqn:E_Decay}, we perform a series of simulations for a range of potential widths $\lambda$ and several values of $\mu$ in both attractive ($\Phi_0 < 0$) and repulsive ($\Phi_0 > 0$) regimes. We extract the coordinates of the turning points to obtain the time-dependent amplitude of the oscillations and plot the fractional amplitude reduction $1 - [\sigma(\tau)/\sigma_0]^6$ vs. $\tau$ in Fig.~\ref{fig:Power_Law}. Based on the scaling argument above, the trajectory should have slope $+1$ on a log-log plot. 

Panels (a)-(c) in Fig.~\ref{fig:Power_Law} demonstrate the scaling for the attractive interaction for a set of parameters $\lambda$, $\mu$, and $\Phi_0 < 0$. 
In addition to the turning points extracted from the simulations, we plot fits $1-[\sigma(\tau)/\sigma_0]^6 = C\tau(\Phi_0\lambda)^2/\mu$ for $C = 5\times 10^{-9}$. 
The fact that the same constant $C$ yields good fits for a range of parameters supports our scaling argument. 
We see that for large $|\Phi_0|$ and $\lambda$ the offset of the fits becomes worse.
This deviation can be explained by the fact that wider (larger $\lambda$) and stronger (larger $|\Phi_0|$) potentials lead to longer interaction times $\tau_\mathrm{int}$. 
Consequently, the approximations used in deriving the scaling law become less appropriate.
We obtain similar results for the repulsive interaction, as shown in Fig.~\ref{fig:Power_Law}(d)-(f). 
For the repulsive potential, the best-fit value of the constant is larger, $C=7\times 10^{-9}$.

\begin{figure}[htb]
    \centering
    \includegraphics[width=\columnwidth]{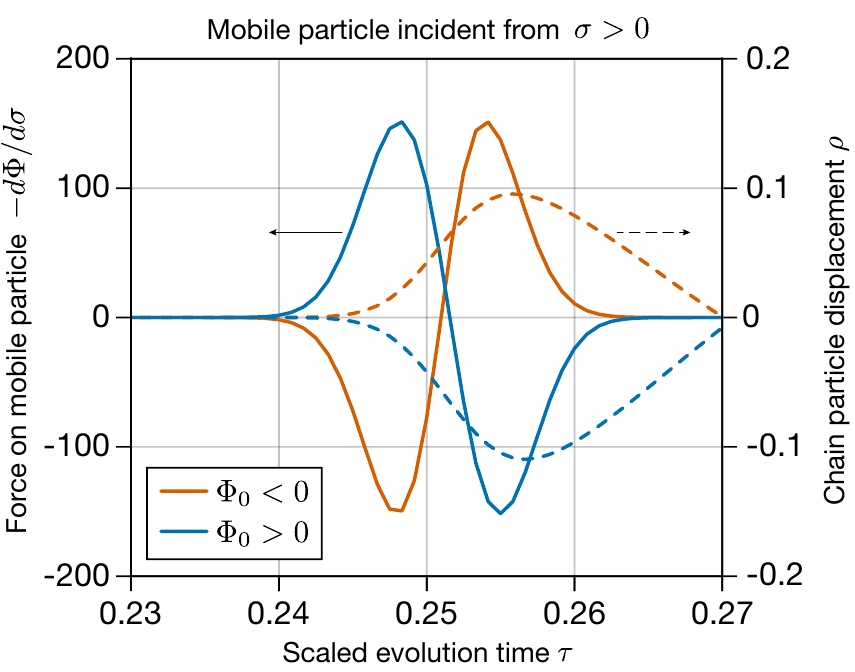}
    \caption{\textbf{Difference between attractive and repulsive interactions.} Attractive and repulsive interactions with the same strength $|\Phi_0|$ and length scale $\lambda$ lead to slightly different dynamics. A mobile particle incident from $\sigma>0$ experiences a force with nearly identical profile (with opposite amplitude), but the displacement of the chain particle is slightly larger for repulsive interactions, and the duration is slightly longer. This extra displacement leads to slightly faster dissipation, as seen in Fig.~\ref{fig:Power_Law}.
    $\omega_\text{min}=2$, $\omega_\text{max}=20$, $\mu = 2$, $\lambda=2$, $\Phi_0 = \pm500$.}
    \label{fig:Attractive_repulsive}
\end{figure}

\begin{figure*}
    \centering
    \includegraphics[width=\textwidth]{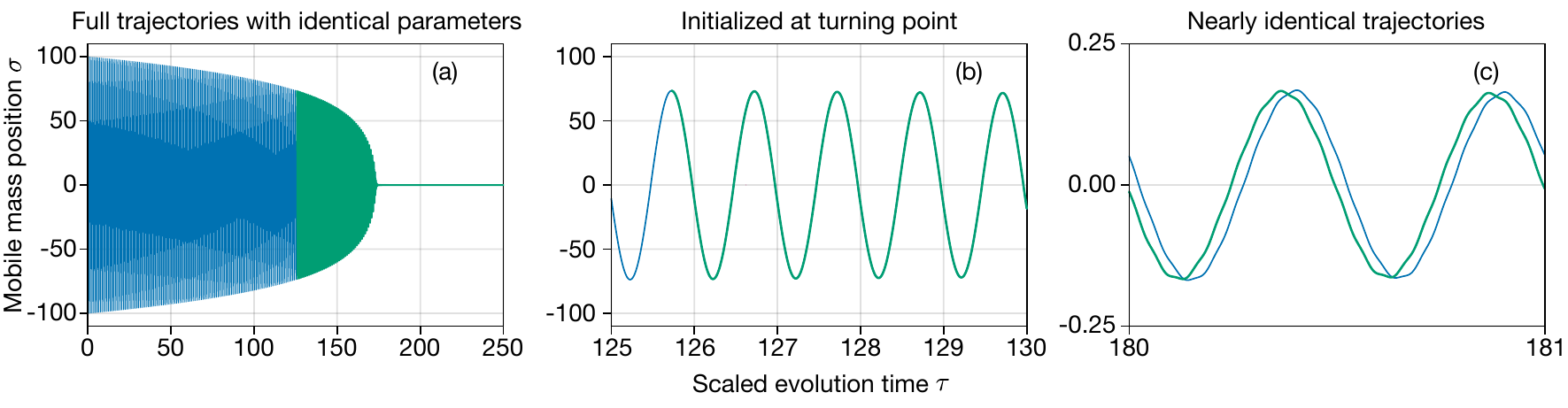}
    \caption{\textbf{The role of history in dissipation.} A comparison of two trajectories for $\omega_\text{min}=2, \omega_\text{max}=20$, $\mu=2$, $\lambda = 4$, and $\Phi_0 = -500$. The blue trajectory starts at $\sigma_0 = 100$ at $\tau=0$ and the green one at $\sigma_0 \approx 73.5$ at $\tau\approx126$. (a) An overlay of the two trajectories shows an identical envelope, magnified for a few oscillations in (b). (c) At later times, we see a mild disagreement in the dissipation-free regime.}
    \label{fig:Different_Starts}
\end{figure*}
\begin{figure}
    \centering
    \includegraphics[width=\columnwidth]{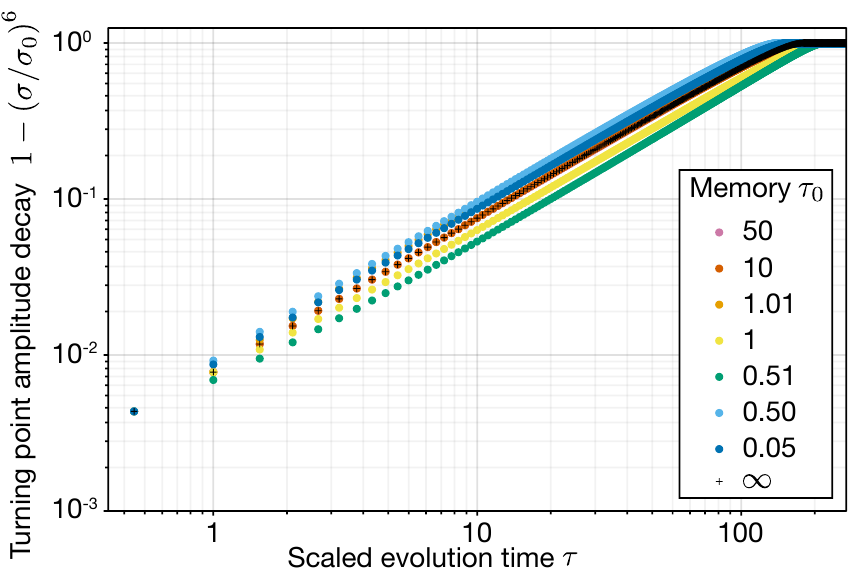}
    \caption{\textbf{The role of memory in dissipation.} The amplitude reduction of the mobile particle at its turning points for $\omega_\text{min}=2, \omega_\text{max}=20$, $\mu=2$, $\lambda = 4$, and $\Phi_0 = -500$ shows a relatively weak dependence on $\tau_0$.}
    \label{fig:Memory_SingleT0}
\end{figure}

The different values of $C$ arise from the difference in effective interaction time between attractive and repulsive interactions.
As an example, we consider the case of $\lambda = 2$ and $\mu = 2$, corresponding to the yellow data points in Fig.~\ref{fig:Power_Law}(b) and (e).
Looking in detail at the first encounter between the mobile particle and the chain mass, we plot the displacement of the chain particle $\rho$ and $-d\Phi/d\sigma$ for both signs of interaction in Fig.~\ref{fig:Attractive_repulsive}.

The interaction window is fairly short, amounting to about $8\%$ of the mobile particle's travel time between turning points.
Despite this short time, we see that the force and displacement curves are rather smooth, justifying our choice of time step in the calculations discussed in Sec.~\ref{sec:Computational_Procedure}.
Although the shapes of the force profiles that the chain mass trajectories are very similar between the attractive and repulsive interactions, we can see that the time scale is slightly longer and the maximum displacement of the chain mass is larger for the repulsive interaction. 
This difference is explained by the order of acceleration/deceleration that the mobile particle experiences.
If the interaction is attractive, the mobile particle speeds up, being pulled forward by the chain mass, and then slows down as the chain particle `tugs' on it.
Conversely, if the interaction is repulsive, the mobile particle first experiences braking, followed by an acceleration.
Consequently, the mobile particle moves slower when it passes the chain mass in the repulsive configuration than it does in the attractive one, meaning that the effective contact time is larger in the repulsive case.
The difference in the interaction time is directly related to the displacement of the chain particle and, therefore, the amount of elastic energy stored in the chain.

\subsection{Role of Memory}
\label{sec:Role_of_Memory}

Next, we discuss the impact of the memory term $\rho(\tau)$ on dissipative behavior through two calculations.

We start by computing two particle trajectories with identical system parameters as in Fig.~\ref{fig:General_Example}, but where the second particle is initialized at one of the later-time turning points of the first particle.
This is equivalent to erasing the memory of the particle at that turning point.
We plot the computed trajectories in Fig.~\ref{fig:Different_Starts} for $A_1 = 100$ and $A_2\approx 73.5$.
The calculations show that the large-amplitude trajectories for the two simulations coincide quite well. 
Only when the mobile particle falls into the potential well is there a noticeable difference between the two scenarios.

Based on this numerical experiment, we can see that the history of the trajectory in the large-amplitude regime has a minor effect on the motion of the particle.
This result justifies our approach to the dissipation scaling in Sec.~\ref{sec:Dissipation_Scaling}, where we treated each encounter as independent from all the others.
We might be tempted to conclude, based on this result, that the memory time plays a rather minor role in dissipation.

To investigate how dissipation is affected by the memory, we performed a series of calculations using the same parameters from Fig.~\ref{fig:Different_Starts}, but including only the most recent segment of time $\tau_0$. 
That is, we replaced the integral in Eq.~\eqref{eqn:G_1D_unitless} $\int_0^\tau\rightarrow \int^\tau_{\mathrm{max}(0, \tau-\tau_0)}$. 
In Fig.~\ref{fig:Memory_SingleT0}(a), we plot the amplitude decay for several values of the memory time, ranging from $\tau_0=\infty$ to $\tau_0=1/ 20$.
All of the trajectories exhibit nearly the same quasi-power-law scaling, with the largest deviation for $\tau_0=0.51$.
Based on the specific value of $\tau_0$, the dissipation can be either faster or slower than the $\tau=\infty$ case.
Choosing a finite $\tau_0$ leads to small counter-movements in the position of the chain mass as it gradually `forgets' previous interactions.
Due to the decaying behavior of $\Gamma(\tau)$, this behavior is especially prominent for short $\tau_0$.
For specific choices of $\tau_0 \approx 0.5,1$, this motion can lead to pathological behavior as the counter-movements occur during a subsequent interaction with the passing mobile particle.
For certain system parameters, we found the dissipation for pathological $\tau_0$ was qualitatively different from the $\tau_0=\infty$ scaling.
Obviously, a truncated memory kernel is rather artificial.
While an ideal, isolated physical system should have $\tau_0=\infty$, external couplings would lead to faster decay of the memory kernel $\Gamma(\tau)$.
A suitable short memory time shows similar behavior to $\tau_0=\infty$, and also offers the benefit of easier calculation.

\subsection{Multiple Mobile Particles}

We also explore how the presence of multiple non-interacting mobile particles changes their dissipative dynamics.
Initializing the positions of 25 mobile particles starting from rest with a mean position of 100 and standard deviation of 20, we tracked their resulting motion (see Appendix \ref{sec:Multiple_Particles_NonThermal} for details).
We found that each particle shows roughly the same quasi-power-law dissipation, with its timescale increased by the number of other particles `tethered' to the chain mass.
For both attractive and repulsive interactions, the collection of mobile particles and the chain mass eventually exhibit persistent oscillations with frequencies just outside the phonon band, just as in the single-particle case.

\section{Thermalization}
\label{sec:Thermalization}

Having developed a good understanding of dissipation in our model system, we reintroduce the homogeneous thermal motion term in Eq.~\eqref{eqn:r_N_unitless}.
To justify our semi-classical approach, we first verify that the vacuum fluctuations of the chain mass are much smaller than the characteristic interaction length $\lambda$:

\begin{align}
    \langle \rho^2\rangle_{T = 0}
    &= 
    \frac{1}{\mu\pi}\int_{\omega_\mathrm{min}}^{\omega_\mathrm{max}}
    \frac{dz}{\sqrt{z^2 - \omega^2_\mathrm{min}}\sqrt{\omega^2_\mathrm{max} - z^2}}
    \nonumber
    \\
    &= 
    \frac{1}{\mu\pi}\frac{\mathcal{K}\left(1 - \frac{\omega_\mathrm{min}^2}{\omega_\mathrm{max}^2}\right)}{\omega_\mathrm{max}}
    \approx
    \frac{1}{\mu\pi}\frac{\ln\left(\frac{4\omega_\mathrm{max}}{\omega_\mathrm{min}}\right)}{\omega_\mathrm{max}}\,,
    \label{eqn:rho_avg}
\end{align}
where $\mathcal{K}(x)$ is the complete elliptic integral of the first kind.
For the system parameters used above, 
$\mu=2,\omega_\text{min}=2,\omega_\text{max}=20$, we find $\sqrt{\langle \rho^2\rangle}_{T = 0}\approx 0.17$.
Lengths in our simulations are scaled by $l_M$, and the potential width $\lambda = 4$, so the scale of vacuum fluctuations is negligible.

\subsection{Single Particle}
\label{sec:Single_Particle}

\begin{figure*}
    \centering
    \includegraphics[width=\textwidth]{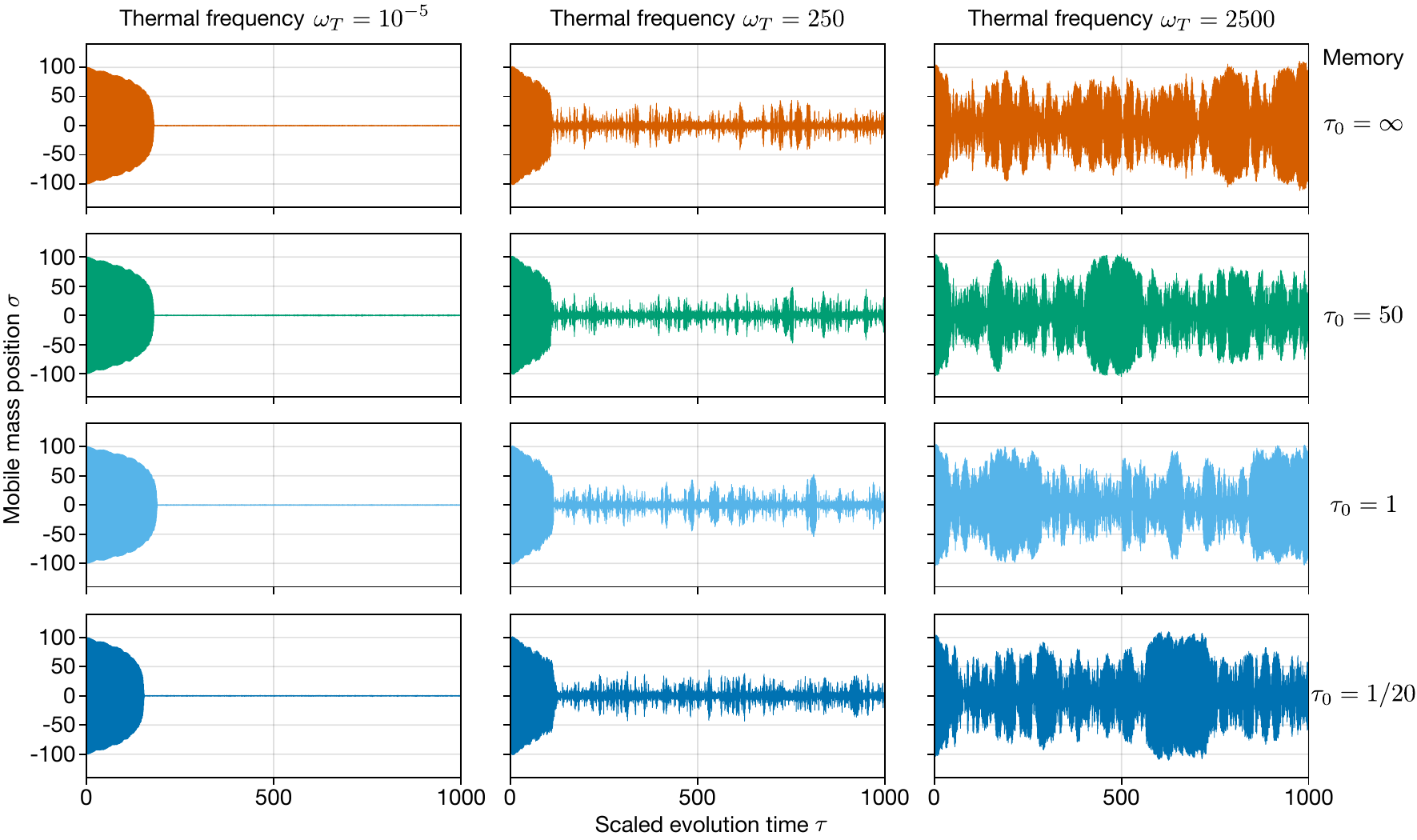}
    \caption{\textbf{General example with thermal motion included.} Trajectories of a single mobile particle $\sigma(\tau)$ for $\omega_\text{min}=2$, $\omega_\text{max}=20$, $\mu = 2$, $\lambda = 4$, $\Phi_0 = -500$ with $\sigma_0 = 100$ for three temperatures $\omega_T$: $10^{-5}$ (left column), $250$, (middle column), $2500$ (right column) reveal dissipation and fluctuation. The four rows correspond to different memory times, from top to bottom: $\infty$, $50$, $1$, and $1/20$.}
    \label{fig:Thermal_Example}
\end{figure*}

We start by computing individual particle trajectories with the same system parameters as in Fig.~\ref{fig:General_Example}, but now including the $\rho^H(\tau)$ term in Eq.~\eqref{eqn:EOM_discrete}, calculated following the procedure given in Sec.~\ref{sec:Computational_Procedure}.
The results are shown in Fig.~\ref{fig:Thermal_Example} for three different temperatures ($\omega_T = k_B T/ E_M = 10^{-5}$, 250, and 2500, arranged in columns) and four different memory times ($\tau_0=\infty$, $50$, $1$, and $1/20$, arranged in rows).
For a given temperature, the simulations with different memory times have identical $\rho^H(\tau)$.

As expected, increasing the temperature of the chain (going left to right in Fig.~\ref{fig:Thermal_Example}) produces a larger amplitude of the mobile particle motion at later times. 
Although changing the memory time does alter the trajectory, the amplitude of the mobile particle motion remains nearly the same. 
Qualitatively, the trajectories for the particles at late times seem to only depend on the chain temperature.
To quantitatively verify that these systems exhibit fluctuation behavior, we will study the statistical properties of the trajectory.
We expect the mobile particle's energy distribution to be determined by the chain temperature. 
In particular, it should follow the Boltzmann distribution $P(E)\propto \exp\left(-\frac{E}{\hbar\Omega_T}\right)$, or $P(\mathcal{E})\propto \exp\left(-\mathcal{E}/\omega_T\right)$, where $\mathcal{E}$ is the total energy $E$ of the mobile particle in units of $E_M$.

We repeat the calculations in Fig.~\ref{fig:Thermal_Example} for a larger set of $\omega_T$'s: 100, 250, 500, 1000, and 2500 with the same four memory times using $\Phi_0 = \pm 500$.
After calculating the trajectories, we extract the total energy $\mathcal E$ for the particle at each of the $1.2\times 10^6$ time steps. 
In our dimensionless formulation, potential energy (in units of $E_M$) is given by $\sigma_{j,\alpha}^2/2 + \Phi\left(\rho_\alpha, \sigma_{j,\alpha}\right)$, while the kinetic energy is $[(\sigma_{j,\alpha + 1} - \sigma_{j,\alpha})/(2\pi \delta)]^2/2$.
To eliminate the effects of the dissipative portion of the trajectory, we drop the first $10^5$ steps for the repulsive potential and $1.5\times 10^5$ for the attractive one (as we observed, the repulsive potential exhibits faster dissipation). 
Finally, we divide the energies by $\omega_T$ and build a normalized histogram to extract the probability distribution $P(\mathcal{E})$ as a function of $\mathcal{E} / \omega_T$.
On a plot of $\ln \left[P(\mathcal E)\right]$ vs.~$\mathcal{E}/\omega_T$, a particle at thermal equilibrium will have slope $-1$.

\begin{figure*}
    \centering
    \includegraphics[width = \textwidth]{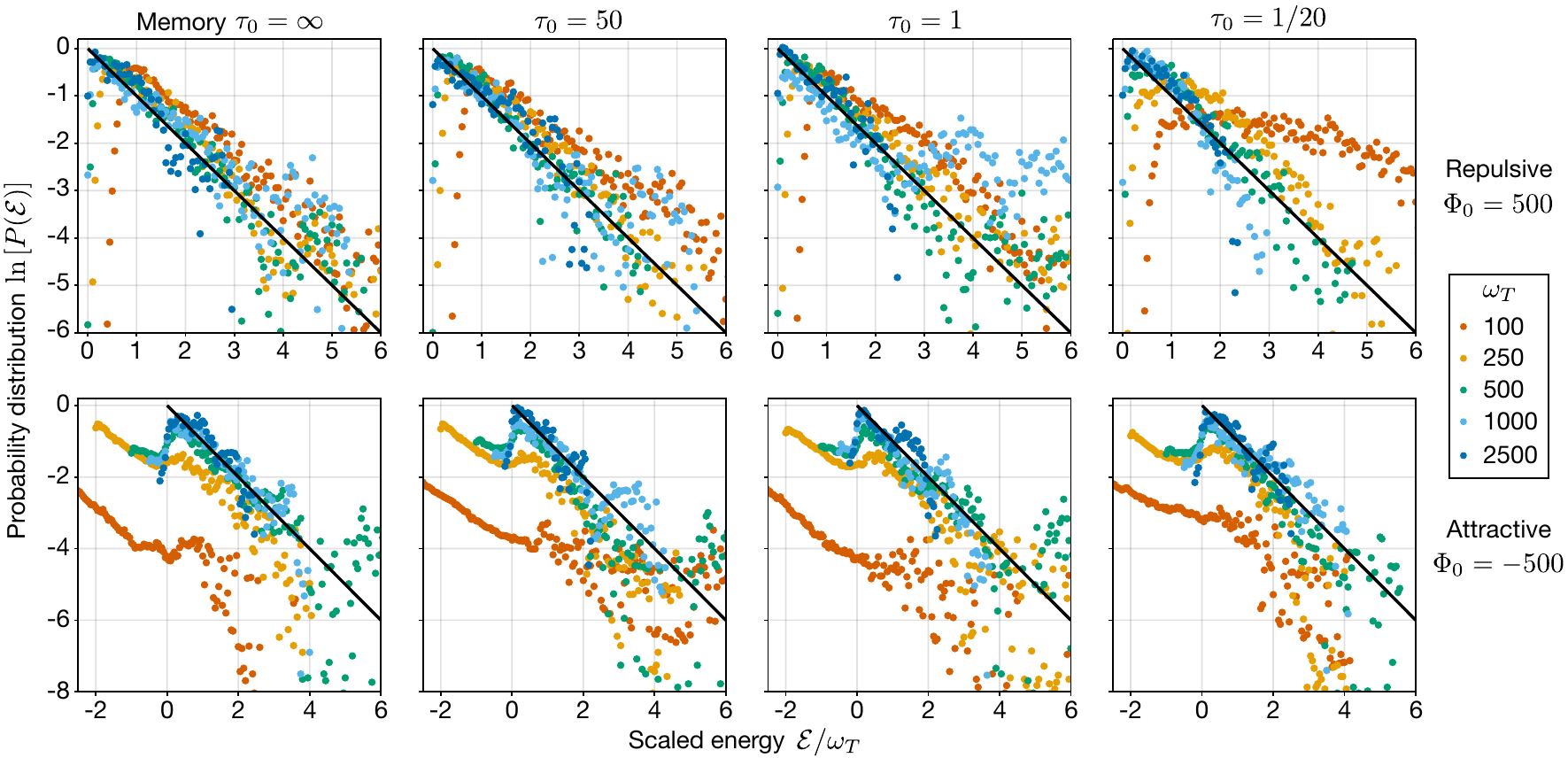}
    \caption{\textbf{Energy probability distribution.} Logarithm of the probability distribution of total energy per particle $\mathcal E$ vs. the total energy divided by the chain temperature $\omega_T$ for $\omega_\mathrm{min} = 2$, $\omega_\mathrm{max} = 20$, $\mu = 2$, $\lambda = 4$, and $\Phi_0 = \pm500$. The top (bottom) row contains the results for the repulsive (attractive) interactions. The four columns correspond, from left to right, to $\tau_0 = \infty$, $50$, $1$, and $1/20$. The black lines each have a slope of $-1$, corresponding to the Boltzmann distribution $P = e^{-\frac{\mathcal{E}}{\omega_T}}$. The collapse of the data points onto the black curve indicates that the mobile particle follows the Boltzmann distribution with the same temperature as the chain.}
    \label{fig:Single_Energy}
\end{figure*}

For the calculations in the top row of Fig.~\ref{fig:Single_Energy} (with repulsive interaction), we observe that the probability distributions mostly collapse onto a common line with a $-1$ slope, as expected.
The deviation at low energies is due to the repulsive interaction between the chain and the mobile particle, so that the latter never has zero energy.
For larger $\omega_T$, the time between interactions with the chain mass is longer, so our finite simulations undersample the rarer high-energy time steps.
Unlike the dissipative case, a short memory time $\tau_0$ can also give rise to pathological behavior, particularly for lower temperatures.
In this case, the interactions between the mobile particle and chain mass become more frequent and take longer, so truncating the memory leads to problematic counter-motion.

In the case of attractive interactions, shown in the bottom row of Fig.~\ref{fig:Single_Energy}, we see that for sufficiently high $\omega_T$'s, we obtain the expected linear relationship between $\ln\left[P(\mathcal{E})\right]$ and $\mathcal{E} / \omega_T$. 
For lower $\omega_T$'s, however, there is a qualitatively different behavior for negative energies.
In this regime, the mobile particle spends a significant fraction of time in very close proximity to the chain mass, essentially trapped in the interaction potential.
Consequently, the mobile particle becomes tethered to the chain particle, and we do not expect it to exhibit the proper statistics.

For the repulsive case, where the mobile particle does not get as tightly tethered, another pathology arises.
For low temperatures $\omega_T \lesssim \Phi_0$, the mobile particle will spend a large fraction of time near the chain mass.
In this regime, the pathological counter-motion produced by truncating the memory integral has a greater impact on the mobile particle's motion.
Thus, shorter $\tau_0$ can lead to anomalously high apparent temperatures, as seen in the top right panel of Fig.~\ref{fig:Single_Energy}.

\subsection{Multiple Particles}
\label{sec:Multiple_Particles}

To improve the statistics while also making the system more realistic, we also performed calculations with 25 mobile particles, with identical system parameters as in Fig.~\ref{fig:Single_Energy}.
We initialized the particles starting from rest, positioned according to a normal distribution with a mean of 100 and standard deviation of 20.
To make the comparison between different simulation runs more robust, we used the same starting positions by employing the same random seed.

\begin{figure}
    \centering
    \includegraphics[width = \columnwidth]{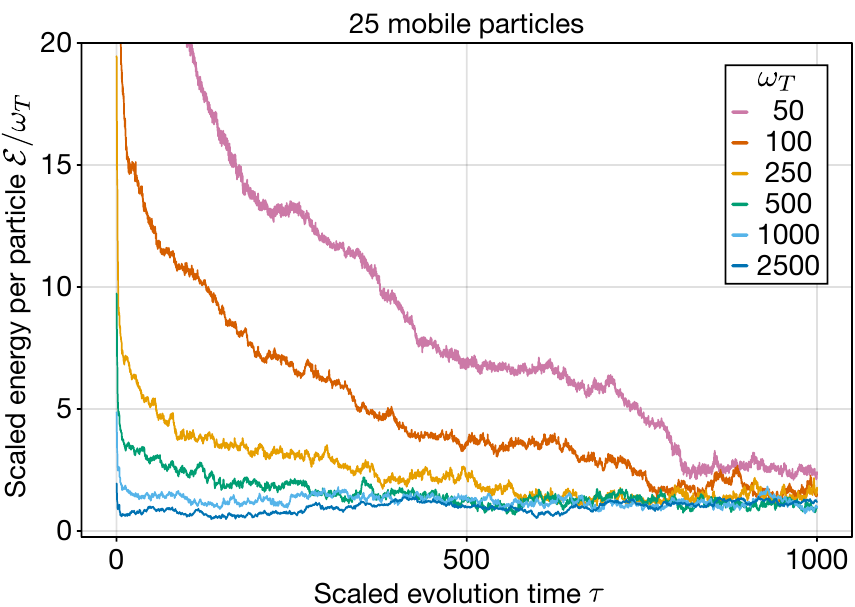}
    \caption{\textbf{Thermal equilibration of mobile particles.} Average energy per mobile particle in an ensemble of 25 mobile particles for $\omega_\mathrm{min} = 2$, $\omega_\mathrm{max} = 20$, $\mu = 2$, $\lambda = 4$, and $\Phi_0 = 500$. The trajectories approach $\mathcal{E} / \omega_T = 1$ as the temperature of the particle ensemble approaches $\omega_T$.}
    \label{fig:Energy_over_time}
\end{figure}

To illustrate the rate of thermal equilibration, we plot the average energy per particle as a function of time for a repulsive potential in Fig.~\ref{fig:Energy_over_time}.
Except in the case of $\omega_T=50$, where the repulsive interaction potential substantially shifts the minimum attainable energy of the mobile particles, each of the ensembles eventually approaches thermal equilibrium, $\mathcal{E}\rightarrow \omega_T$.
We see that as $\omega_T$ increases, the equilibration time decreases substantially.
Intuitively, this pattern makes sense, as the mobile particles dissipate more energy in the case of lower $\omega_T$.
We do not show results of similar calculations with an attractive potential 
because once the particles become tethered to the chain mass, their negative energy significantly skews the average.


%

\begin{figure}
    \centering
    \includegraphics[width = \columnwidth]{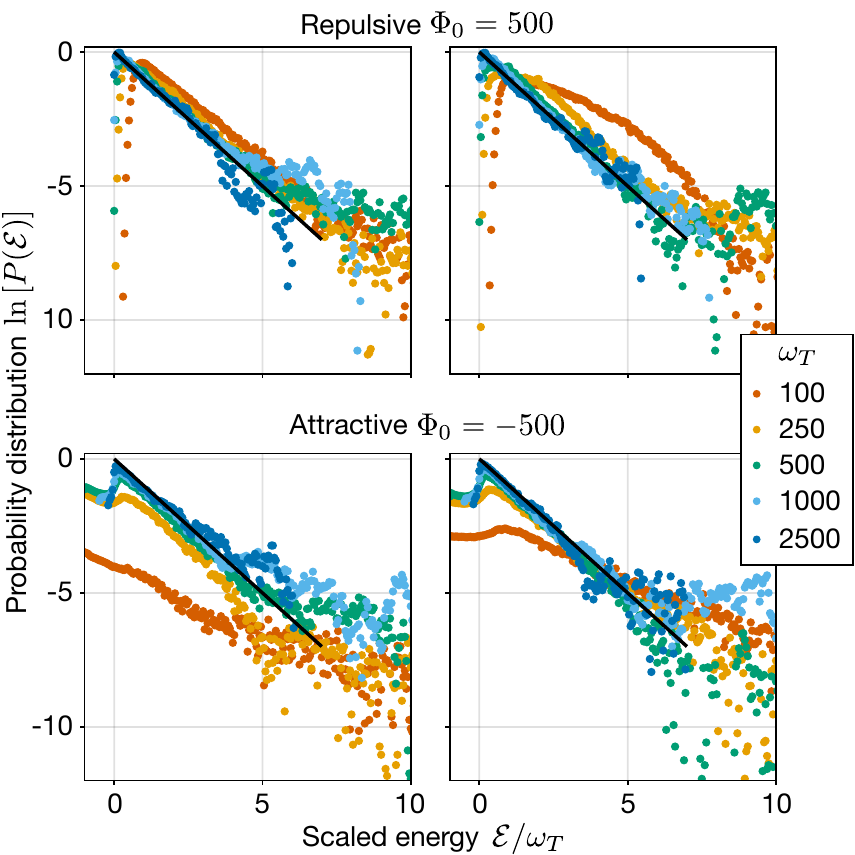}
    \caption{\textbf{Energy probability distribution for an ensemble.} Logarithm of the probability distribution of total energy per particle vs. the total energy per particle divided by the chain temperature $\omega_T$ for $\omega_\mathrm{min} = 2$, $\omega_\mathrm{max} = 20$, $\mu = 2$, $\lambda = 4$, and $\Phi_0 = \pm500$ obtained from a simulation with 25 mobile particles. The top (bottom) row contains the results for the repulsive (attractive) interaction. The left column corresponds to $\tau_0 = \infty$, while the right one is $\tau_0 = 1/20$. The black lines have slopes of $-1$, corresponding to the Boltzmann distribution $P = e^{-\frac{\mathcal{E}}{\omega_T}}$.}
    \label{fig:Energy_Distribution}
\end{figure}

Next, we construct the probability distribution of the total energy per particle, similar to Fig.~\ref{fig:Single_Energy}, but for just two memory times $\tau_0=\infty, 1/20$. 
We expect that the addition of multiple mobile particles will reduce the effect of anomalous counter-motion caused by truncating the memory.
In the single-particle case, the counter-motion was especially problematic because it tended to reverse the energy transfer of the previous interaction.
In the many-particle case, however, the numerous intervening interactions tend to overwhelm the counter-motion as the response to 24 other particles dwarfs the counter-motion originating from a single contact.
Put differently, from the point of view of a given particle, additional interactions conceal the counter-motion by effectively adding randomness to the chain mass' trajectory.
We see from the results in Fig.~\ref{fig:Energy_Distribution} that the 25-fold increase in the number of data points improves the agreement with the Boltzmann distribution.
We also observe that reducing the memory from infinity to $1/20$ does not substantially change the resulting distribution, except for at low temperatures.

Although we did not include any interaction between the mobile particles, they do, in fact interact indirectly via the chain. To ensure that this interaction does not impact the individual particle statistics by giving rise to collective behavior, we found that the cross-correlations of the particle positions are rather small (see Appendix~\ref{sec:Correlation}).

\begin{figure}
    \centering
    \includegraphics[width=\columnwidth]{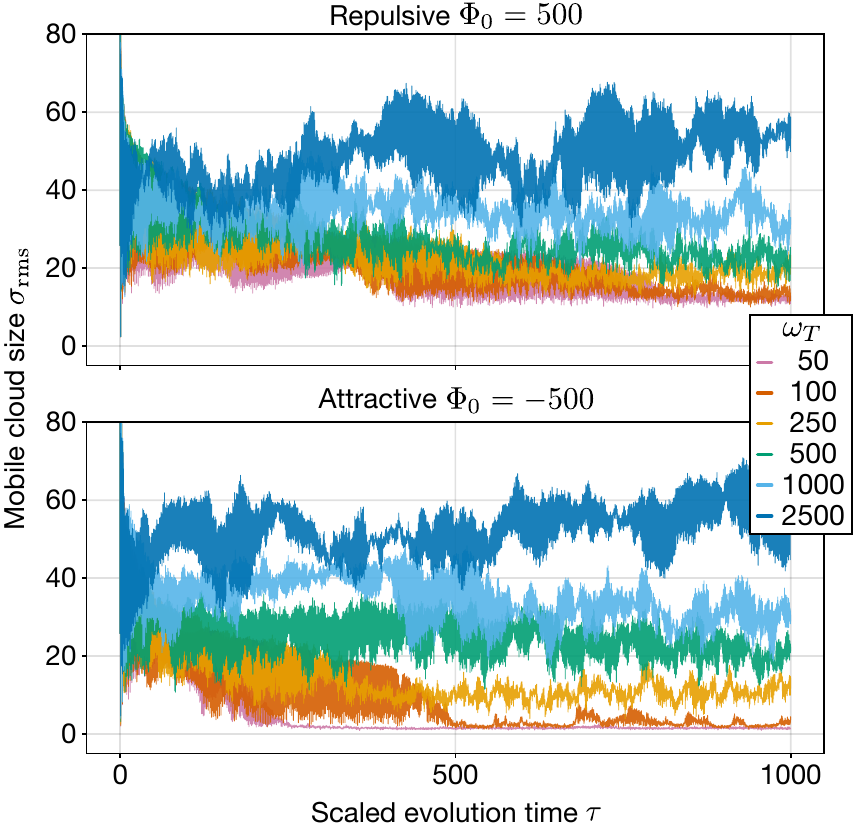}
    \caption{\textbf{Cloud size}. Root mean square displacement of the mobile particles for repulsive (top) and attractive (bottom) interactions with the chain. The data is obtained from the simulations used to produce the left column of Fig.~\ref{fig:Energy_Distribution}. The cloud size within the harmonic trap roughly matches the chain thermal energy scale, $\sigma_\text{rms}^2\approx\omega_T$.}
    \label{fig:Cloud_Size}
\end{figure}
\begin{figure}
    \centering
    \includegraphics[width=\columnwidth]{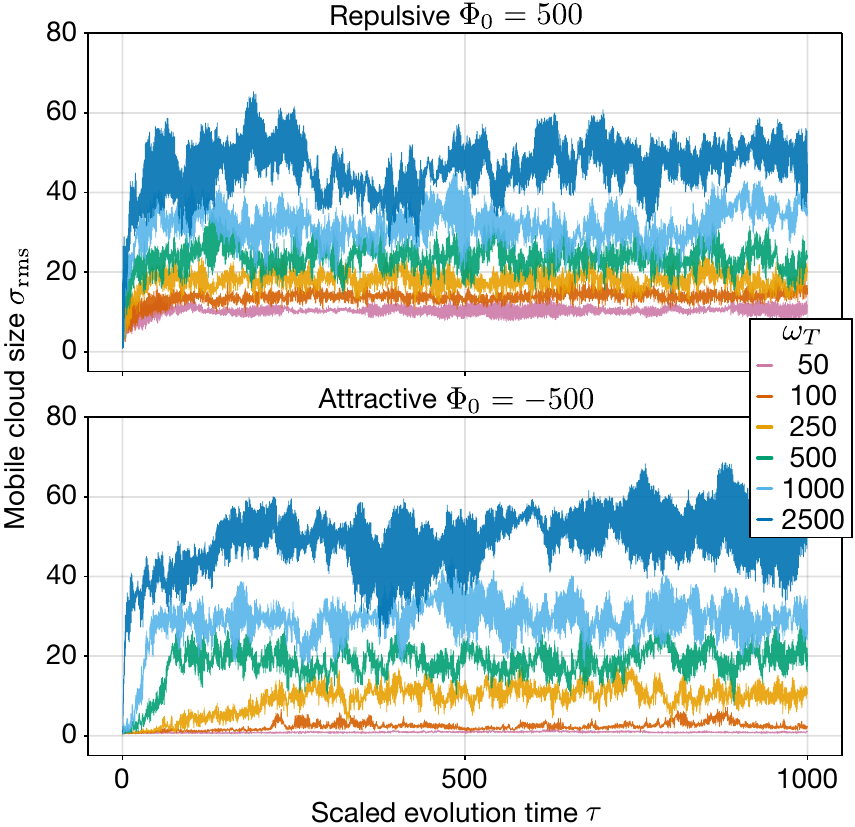}
    \caption{\textbf{Cloud size increase}. Root mean square displacement of the mobile particles for the repulsive (top) and attractive (bottom) interaction with the chain. The system parameters are the same as in Fig.~\ref{fig:Cloud_Size}, but now the particles are initialized close to the bottom of the potential well. The final cloud size for each temperature and interaction agrees with the corresponding result from Fig.~\ref{fig:Cloud_Size} indicating that the initial conditions do not play a role in determining the spread of the mobile particles. Again, the final cloud size within the harmonic trap roughly matches the chain thermal energy scale, $\sigma_\text{rms}^2\approx\omega_T$.}
    \label{fig:WarmingUp_Cloud_Size}
\end{figure}

Naturally, raising the temperature of the system increases the oscillation amplitude of the mobile particles, leading to a larger cloud size.
In a harmonic trap, the cloud size can be used as a proxy for the temperature.
To illustrate this increase, we plot the root-mean-squared displacement $\sigma_\mathrm{rms}$ for a range of temperatures for both attractive and repulsive interactions in Fig.~\ref{fig:Cloud_Size}.
We also investigate cloud heating by initializing the 25 particles normally distributed around the origin with a standard deviation of 1 and tracking the increase of $\sigma_\mathrm{rms}$ over time in Fig.~\ref{fig:WarmingUp_Cloud_Size}.
We see that the final $\sigma_\mathrm{rms}$ for different temperatures is the same regardless of the initial positions of the mobile particles.

\subsection{No Memory}
\label{sec:No_Memory}

\begin{figure}
    \centering
    \includegraphics[width=\columnwidth]{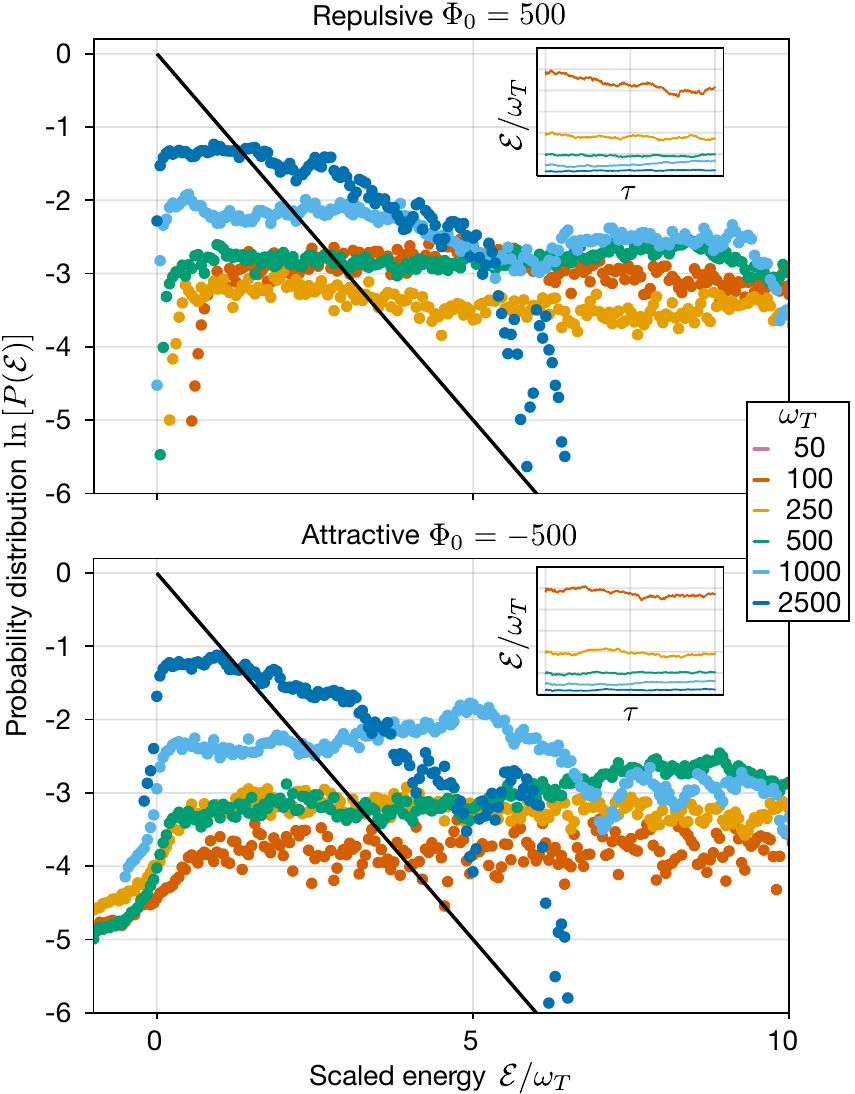}
    \caption{\textbf{Simulations without memory.} 
    Logarithm of the probability distribution of total energy per particle vs. the total energy per particle divided by the chain temperature $\omega_T$ for the same parameters as in Fig.~\ref{fig:Energy_Distribution} but with $\tau_0 = 0$. The insets show the evolution of energy per particle with time. The lack of data collapse onto the Boltzmann distribution indicates the absence of thermalization.}
    \label{fig:Zero_Mem}
\end{figure}

Based on our simulations, we see thermalization for a wide range of memory times $\tau_0$, even in the Markovian limit at $\tau_0\rightarrow 0$.
Computationally, it is much simpler to eliminate the recoil term altogether, setting $\rho(\tau)=\rho^H(\tau)$.
However, we find that in the absence of a recoil term, the mobile particles do not approach a Boltzmann distribution of energy, even after very long times, as seen in Fig.~\ref{fig:Zero_Mem}.
In these simulations, the average energy per particle remains roughly constant, as seen in the insets, in contrast to the dissipation seen in Fig.~\ref{fig:Energy_over_time}.
In practical terms, any experimental implementation of such a model requires some form of feedback to display thermalization.

\section{Proposed Implementations}
\label{sec:Implementations}

Several experimental platforms for investigating 1D physics have arisen in the past few decades.
Here we outline some candidate systems that could be used as realizations to validate the simple model we have developed.
We discuss some of the technical hurdles of each platform, as well as possible avenues for extension.

\subsection{Trapped ion and dimple potential}
\label{ssec:IonsDimple}

One of the most common trapping geometries for ions, the linear Paul trap, confines ions to motion in a 1D harmonic potential \cite{bylinskii_tuning_2015}.
Another geometry using surface electrodes is emerging as a platform for quantum computing, and can be engineered to generate tunable potentials \cite{romaszko_engineering_2020} or to couple to light \cite{ivory_integrated_2021}.
In both trap geometries, it is possible to load a small number of ions, or even a single ion, into the trap.
These trapped ions take the role of the mobile particle in our model, since they move axially in a quasi-1D harmonic potential.
Typical axial trap frequencies vary from several kHz \cite{west_tunable_2021} to the MHz regime \cite{bruzewicz_measurement_2015}, with motional heating rates as low as a few quanta/s at cryogenic temperatures \cite{chiaverini_insensitivity_2014}.
In order to implement the effective bath coupling, the trap potential near the center could be modified using a small dc electrode or a tightly-focused laser beam, which would produce a dimple potential~\cite{he_ion_2021, romaszko_engineering_2020}.

By monitoring the position of the ion using a weak probe beam \cite{sames_continuous_2018, bushev_shot-noise-limited_2013} or fluorescent light \cite{cerchiari_measuring_2021}, the position and depth of the dimple could be modified with a feedback loop according to the dynamics described in Eq.~\eqref{eqn:r_N_unitless}.
Similar feedback loops using large-scale electric fields produced by trap electrodes have been implemented in two main ways: cold damping \cite{bushev_feedback_2006} and parametric cooling \cite{sames_continuous_2018}.

In this implementation, adding multiple mobile ions to the trap is straightforward.
However, the Coulomb coupling between mobile ions is much stronger than the effective bath coupling; we will explore this strongly-interacting case in future work.
In contrast to other cooling methods, our scheme only perturbs the trap locally.
It could thus potentially be useful in contexts where the trap frequency needs to remain stable, such as in ion-based sensing.






\subsection{Hybrid atom-ion system}
\label{ssec:IonsParticle}

Another experimental system that has been extensively studied recently consists of a small number of ions immersed in a cloud of neutral atoms \cite{harter_cold_2014, tomza_cold_2019}.
These hybrid systems are useful for studying collision dynamics, or to use the ions as sensitive probes of the cloud \cite{veit_pulsed_2021, cetina_ultrafast_2016}.
In a typical implementation, an ion is confined using a Paul trap, with neutral atoms overlapped using optical potentials \cite{schmid_dynamics_2010}.
However, since the radiofrequency potential for the ion is much deeper than the optical trap experienced by the neutral atoms, the length scale for ion motion is smaller than the typical neutral cloud size.
Recent experiments have employed optical traps for both the ion and the neutral cloud \cite{weckesser_observation_2021, schmidt_optical_2020, weckesser_trapping_2021}.
An ion confined to quasi-1D motion in an optical dipole trap that intersects a cloud of neutral atoms that are confined in a separate optical dipole trap could realize our model.
The large number of degrees of freedom in the neutral cloud would let it serve as an effective bath that couples to the ion's secular motion via collisions at the trap center, where the neutral cloud's size and density would determine the parameters of the coupling $\lambda, \Phi_0$.
The initial kinetic energy of such an ion is on the order of 100~\textmu K$/k_B$, and the temperature of a typical ultracold cloud is $\sim 1$~\textmu K.

Using a similar setup, but in another regime, where the ion undergoes many collisions with the neutral atoms, a similar damping effect occurs, known as buffer gas cooling \cite{devoe_power-law_2009, meir_dynamics_2016-1, feldker_buffer_2020}.

One potential pitfall of this approach is that the ac Stark shift induced by the neutral cloud's trapping light will induce a static dimple in the ion's potential.
This additional dimple could enhance or slow down the dissipation rate, depending on details of the trap laser.
One way to circumvent this issue would be to trap the neutral cloud using a laser at a magic wavelength of the ion \cite{kaur_magic_2015}.

Similar to the ion/dimple approach, this platform can easily scale to multiple mobile particles, and the properties of the bath coupling can be tuned.
For example, the neutral atom cloud size and density can be varied by changing the optical trap power, and its temperature can extend from the thermal to quantum regimes.
We will explore the implications of a quantum bath in future work.


\subsection{Neutral atom bright soliton and dimple}
\label{ssec:Soliton}

Another quasi-1D system with slow dynamics is bright solitons of bosonic neutral atoms in optical waveguides. 
Bright solitons have been implemented with several atomic species \cite{khaykovich_formation_2002, strecker_bright_2003, cornish_formation_2006, lepoutre_production_2016, meznarsic_cesium_2019}, and are commonly trapped in a far-detuned optical dipole potential which provides harmonic confinement.
The lifetime of such solitons can be as high as 3~s, apparently limited by atom loss due to background collisions \cite{strecker_bright_2003}.
Several groups have also introduced optical dimple potentials \cite{mcdonald_bright_2014, nguyen_formation_2017, marchant_controlled_2013, wales_splitting_2020}, where the size of the dimple is usually $\gtrsim 2$~\textmu m (smaller than a typical soliton size of $\sim 10$~\textmu m) and the depth/height of the dimple can be tuned over four orders of magnitude \cite{marchant_quantum_2016}.

The harmonic axial motion can be varied with a combination of magnetic fields and optical potentials from $\sim 5$--$100$~Hz.
With non-destructive techniques, up to 50 images of a cloud can be acquired \cite{seroka_repeated_2019}, allowing many oscillation periods of feedback.

This scheme, due to its long timescale for dynamics, would offer the easiest route toward imaging-based feedback, and would allow the most detailed exploration of feedback involving very short memory $\tau_0$.

\subsection{Neutral atoms in waveguide with dimple}
\label{ssec:waveguide}

A similar implementation involving quasi-1D motion of the mobile particles involves an optical waveguide, but with a large number of neutral atoms moving independently. 
This system has the same general setup as the soliton experiments above, however the dynamics of the mobile particles will be complicated by two extra radial degrees of freedom.
We suspect that a modulated dimple would dissipate energy from the axial motion, and weak interactions between the mobile atoms could lead to dissipation in all three dimensions.
In the experiments described in \cite{raman_dissipationless_2001} and further analyzed in \cite{kiehn_superfluidity_2021}, a sinusoidally modulated dimple beam leads to heating of a BEC. This type of modulation is closely related to Floquet engineering (for a review, see \cite{weitenberg_tailoring_2021}).
With some modifications to the experimental protocol, this system could load a thermal gas of atoms, observe the cloud density through non-destructive imaging, then change the dimple beam position based on Eq.~\eqref{eqn:R_j_unitless}, and measure the resulting distribution of kinetic energies in the cloud through standard time-of-flight imaging.





\subsection{Neutral atoms in an optical lattice with dimple} 
\label{ssec:Cigar}

To restrict the motion of the mobile atoms more closely to 1D, they could be confined using a 2D optical lattice.
Around 20--100 atoms would undergo harmonic motion in each of the $\sim 1000$ 1D tubes, with $\Omega_M$ tunable from $\sim1$--$1000$~Hz.
For a strong enough optical lattice, the energy scale of transverse motion $\sim 100$~kHz could be tuned far above other energy scales in the system.
A light sheet focused very tightly along the axis of the atoms' motion could produce a dimple trap with uniform depth across all of the 1D tubes, yielding many independent realizations of our model system.

One of the major challenges of this approach is implementing the feedback necessary to observe dissipation. Inhomogeneities would lead to slightly different $\Omega_M$ in each of the tubes, precluding effective feedback with a single dimple potential. To make the system more uniform, the optical lattice depth could be tuned using optical techniques \cite{hart_observation_2015}.

\subsection{Neutral atoms and optical cavity}
\label{ssec:cavity}

An experimental tool that could more directly probe the effects of the bath modes in our simulations is an optical cavity. If a multi-mode cavity \cite{kollar_adjustable-length_2015} were aligned transverse to an optical waveguide, the moving atoms would couple to the cavity modes only near the center of the trap. The cavity length could be chosen to tune the energy scale of cavity modes ($\omega_\text{min},\omega_\text{max}$ in our model) relative to the kinetic energy of the mobile particles.

The cavity modes could potentially also be populated in a controlled way to emulate different chain temperatures (or non-equilibrium states). These modes might even be controllable in a feedback loop similar to \cite{kroeger_continuous_2020}.




\section{Conclusion}
\label{sec:Conclusion}

Using a simple model of mobile particles in a harmonic potential coupled to a 1D chain of masses, we have demonstrated both dissipation and fluctuation behaviors using semiclassical numerical simulations.
In the absence of thermal fluctuations, individual particles dissipate energy in a quasi-power-law fashion, with a characteristic time that generally follows a simple scaling based on system parameters.
With multiple non-interacting mobile particles, each of them dissipates energy, with a timescale determined by the number of particles already lying near the chain mass.

Once thermal fluctuations are introduced in the chain motion, we find behavior reminiscent of thermal equilibration.
First, as long as the mobile particles aren't trapped in the interaction potential, they approach a Boltzmann distribution of energy, set by the temperature of the chain.
Next, the characteristic size of a cloud of mobile particles in the harmonic potential eventually matches the thermal energy scale of the chain as well.
For all reasonable choices of the memory time, the system exhibits the same thermalizing behavior.
In contrast, with no memory, the mobile particles' motion is not strongly influenced by the chain, and does not reach a Boltzmann distribution.

We showed that this minimal system exhibits fluctuation and dissipation behavior for a wide range of system parameters.
We suspect that the specific form of the chain-particle interaction has little effect on the resulting dynamics.
However, for some specific interactions, such as a delta-function potential, a different numerical approach will be necessary.
We also suspect that the mode structure of the chain has no qualitative impact on the mobile particle trajectories, and will explore this claim in future work.

We proposed several experimental platforms where this minimal setup could be realized, though each introduces some potential complications.
To model more realistic systems, we will explore the case where the mobile particles interact with one another in future work.
We will also explore how multiple couplings to the chain affect the resulting dynamics.
In this case, the harmonic confinement can be removed, and we expect to observe signatures of diffusion of the mobile particles, in addition to dissipative behavior.
This modification will allow us to explore indirect interactions between mobile particles mediated by the chain.
Additionally, we can explore the appearance of friction and extract the effective friction coefficient.
Another natural extension is to modify the structure of the bulk by either altering the phonon dispersion or by increasing the dimensionality of the system. 
Specifically, the larger number of phonon modes in higher dimensions should lead to faster dephasing, making the memory less important.
In several of the proposed implementations, the scales of the system approach the quantum regime, which will require significant modifications to our approach.


More broadly, this work is a step in developing our understanding of drag and diffusion in solid systems in a non-Brownian regime with nonlinear interactions.
By formulating the problem using microscopic ingredients, we are able to explore the validity of the approximations commonly employed in diffusion problems in solid systems.
From a practical standpoint, improved understanding of dissipative processes in solid materials has a direct impact on our ability to design efficient ionic conductors required for the fabrication of solid state batteries.

\acknowledgments

A.R. acknowledges the National Research Foundation, Prime Minister Office, Singapore, under its Medium Sized Centre Programme and the support by Yale-NUS College (through Grant No. A-0003356-42-00). B.A.O. acknowledges support from Yale-NUS College (through Grant Nos. A-0003356-39-00,  A-0000172-00-00, A-0000155-00-00, and C-607-261-026-001).

A.R. and B.A.O. conceptualized the work; A.R. wrote the code which was used by A.R., A.T. and M.C. to run the simulations; A.R. and B.A.O. analyzed the results, prepared the graphics, and wrote the manuscript.

\appendix

\section{Multi-particle Dissipation}
\label{sec:Multiple_Particles_NonThermal}

\begin{figure*}
    \centering
    \includegraphics[width = \textwidth]{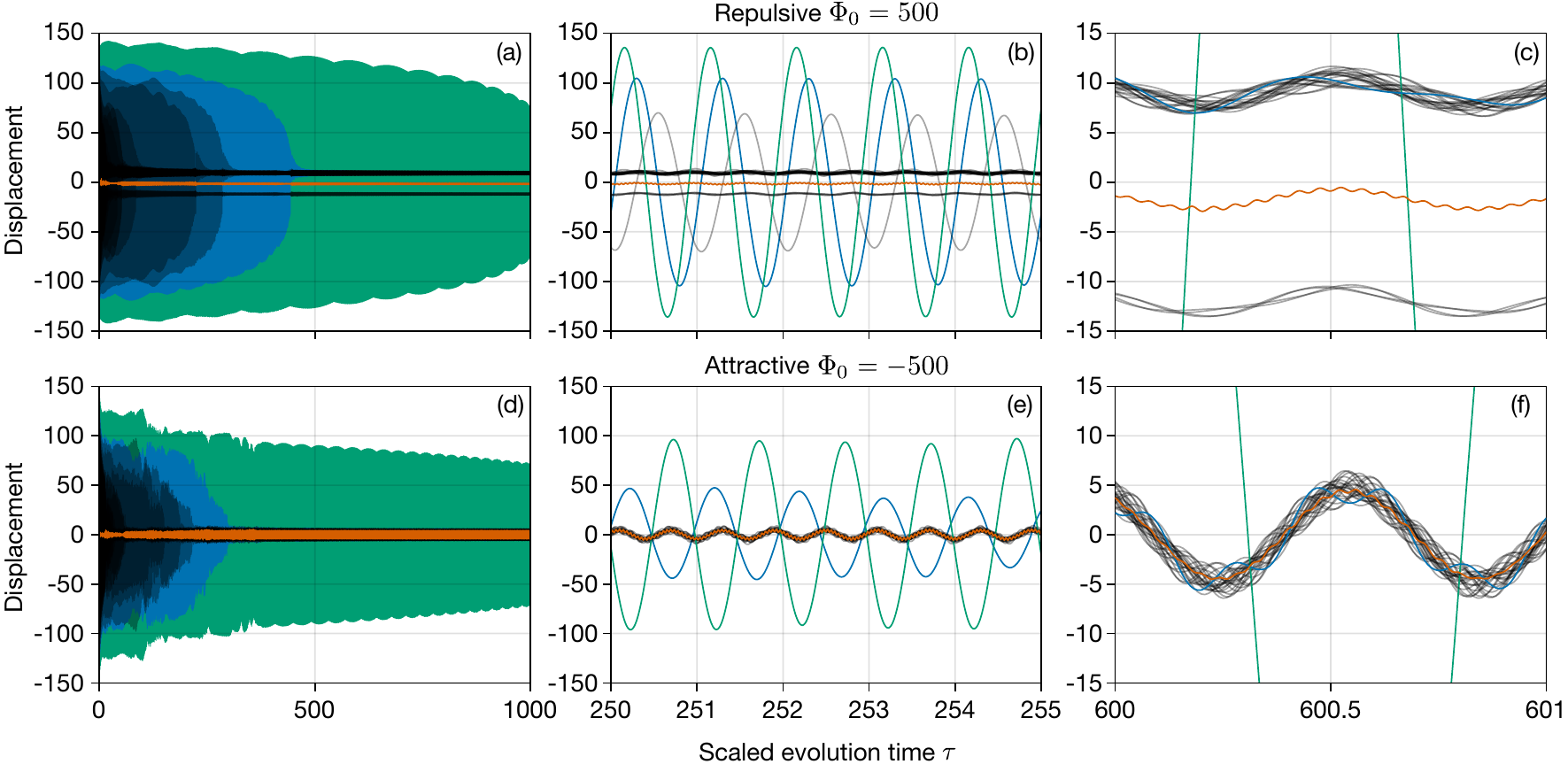}
    \caption{\textbf{Dissipation for a collection of particles.} 25 particles are initialized from rest following a normal distribution with mean of 100 and standard deviation of 20. The system parameters are: $\omega_\mathrm{min} = 2$, $\omega_\mathrm{min} = 20$, $\mu = 2$, $\lambda = 4$, and $\Phi_0 = \pm 500$, with the top (bottom) row corresponding to attractive (repulsive) interaction. The initial positions for the mobile particles are the same for both simulations.}
    \label{fig:Multiple_Particles_NonThermal}
\end{figure*}

To explore how a collection of particles dissipates energy via their interaction with a single chain mass (with no thermal fluctuations), we introduce 25 mobile particles starting from rest with a mean position of 100 and standard deviation of 20.
The trajectories of the mobile particles are plotted simultaneously along with the chain mass for both attractive and repulsive interactions in Fig.~\ref{fig:Multiple_Particles_NonThermal}.

For attractive interactions, similar to the single-particle case, the mobile particles settle near the chain mass, reaching $\left|\sigma_j - \rho\right|\approx 1$ one by one.
Individual particles show qualitatively similar dissipation, roughly following  quasi-power-law trajectories.
After each particle dissipates most of its energy, the collection of masses then oscillates with a persistent amplitude, similar to the behavior seen in Fig.~\ref{fig:General_Example}.

The motion of the chain mass in this regime exhibits two modes which lie further outside the phonon band than the ones observed in Fig.~\ref{fig:General_Example}.
Following the steps leading to Eq.~\eqref{eqn:Fourier}, we obtain that $\rho_\omega = \frac{\Phi_0}{\mu\lambda^2}f_\omega \sum_j\left(\sigma_{j,\omega} - \rho_\omega\right)$ and $\left(1-\omega^2-\frac{\Phi_0}{\lambda^2}\right)\sigma_{j,\omega} + \frac{\Phi_0}{\lambda^2}\rho_\omega = 0$, where $j\in[1,P]$ runs over the mobile particles in the Gaussian well, leading to

\begin{align}
    \begin{pmatrix}
    1-\omega^2-\frac{\Phi_0}{\lambda^2}&0&\dots&\frac{\Phi_0}{\lambda^2}
    \\
    0 & 1-\omega^2-\frac{\Phi_0}{\lambda^2}&\dots&\frac{\Phi_0}{\lambda^2}
    \\
    \vdots&\vdots &\ddots & \vdots
    \\
    -\frac{\Phi_0}{\lambda^2}f_\omega&-\frac{\Phi_0}{\lambda^2}f_\omega&\dots&1 + P\frac{\Phi_0}{\lambda^2}f_\omega
    \end{pmatrix}
    \begin{pmatrix}
    \sigma_{1,\omega}\\\sigma_{2,\omega}\\\vdots\\\rho_\omega
    \end{pmatrix}=0\,.
\end{align}
Taking the determinant of the matrix and setting it equal to zero yields $(P - 1)$ degenerate modes with $\omega = \sqrt{1-\frac{\Phi_0}{\lambda^2}}$ and two more modes with frequencies obtained by solving
\begin{equation}
    (1 - \omega^2)\left(1+P\frac{f_\omega\Phi_0}{\mu \lambda^2}\right) = \frac{\Phi_0}{\lambda^2}\,.
    \label{eqn:small_osc_omega_Multiparticle}
\end{equation}
For the degenerate solutions, the system parameters yield $\omega \approx 5.68$, which lies within the phonon band.
As such, this solution will give rise to decaying oscillations, in accordance with Eq.~\eqref{eqn:Fourier}.
Solving Eq.~\eqref{eqn:small_osc_omega_Multiparticle} for $P = 24$, we get two solutions: one with $\omega \approx 1.65$ and the other with $\omega \approx 25.4$.
This prediction is borne out in our simulations, as seen in Fig.~\ref{fig:Multiple_Particles_NonThermal}(c) by counting the fast and slow oscillations.
As before, because the two frequencies lie outside the phonon band, the energy is not dissipated.
Increasing $P$ further pushes the $\omega$'s away from the phonon band.

For the repulsive case, the situation is somewhat more complex.
Some of the mobile particles settle on the left side of the chain mass, while the rest settle on the right, confined on one side by the chain mass and on the other by the harmonic potential.
The equilibrium positions depend on the exact split of the particles between left and right and determining their locations requires solving transcendental equations.
Moreover, the equivalent of Eq.~\eqref{eqn:Small_Amp_Osc} for the repulsive case is substantially more complicated when multiple particles are involved.
Hence, we do not perform a detailed analysis for the repulsive case.
Instead, we simply demonstrate the presence persistent oscillations for the repulsive case in the bottom row of Fig.~\ref{fig:Multiple_Particles_NonThermal}.

We observe that, unlike the single-particle simulations, the dissipation for the repulsive interaction is not faster than the attractive one.
Given that in the single-particle case the difference in the dissipation rate between the two signs came from a slightly longer effective interaction time for the repulsive interaction, this effect is easily disrupted in the presence of multiple mobile particles.
A closer look reveals that the repulsive interaction actually exhibits slower dissipation.
The reason behind this difference is the uneven deposition of the mobile particles on the two sides of the chain mass.
In the attractive case, the mobile particles that have fallen into the Gaussian well are distributed without any particular order with respect to the chain mass, as can be seen from Fig.~\ref{fig:Multiple_Particles_NonThermal}.
Consequently, they do not substantially impede the minute motion of the chain mass in response to the fast-moving non-tethered particles.
For the repulsive potential, on the other hand, more mobile particles may settle on the positive side of the chain mass than on the negative.
Because of the uneven distribution, the chain mass' energy minimum is slightly to the left of zero.
As a result, the motion of the chain mass is restricted by the mobile masses on one side and by the compressed $\kappa$-spring on the other.
This increased confinement reduces the ability of the chain mass to move, reducing its dissipation ability.

%
\begin{figure*}
    \centering
    \includegraphics[width = \textwidth]{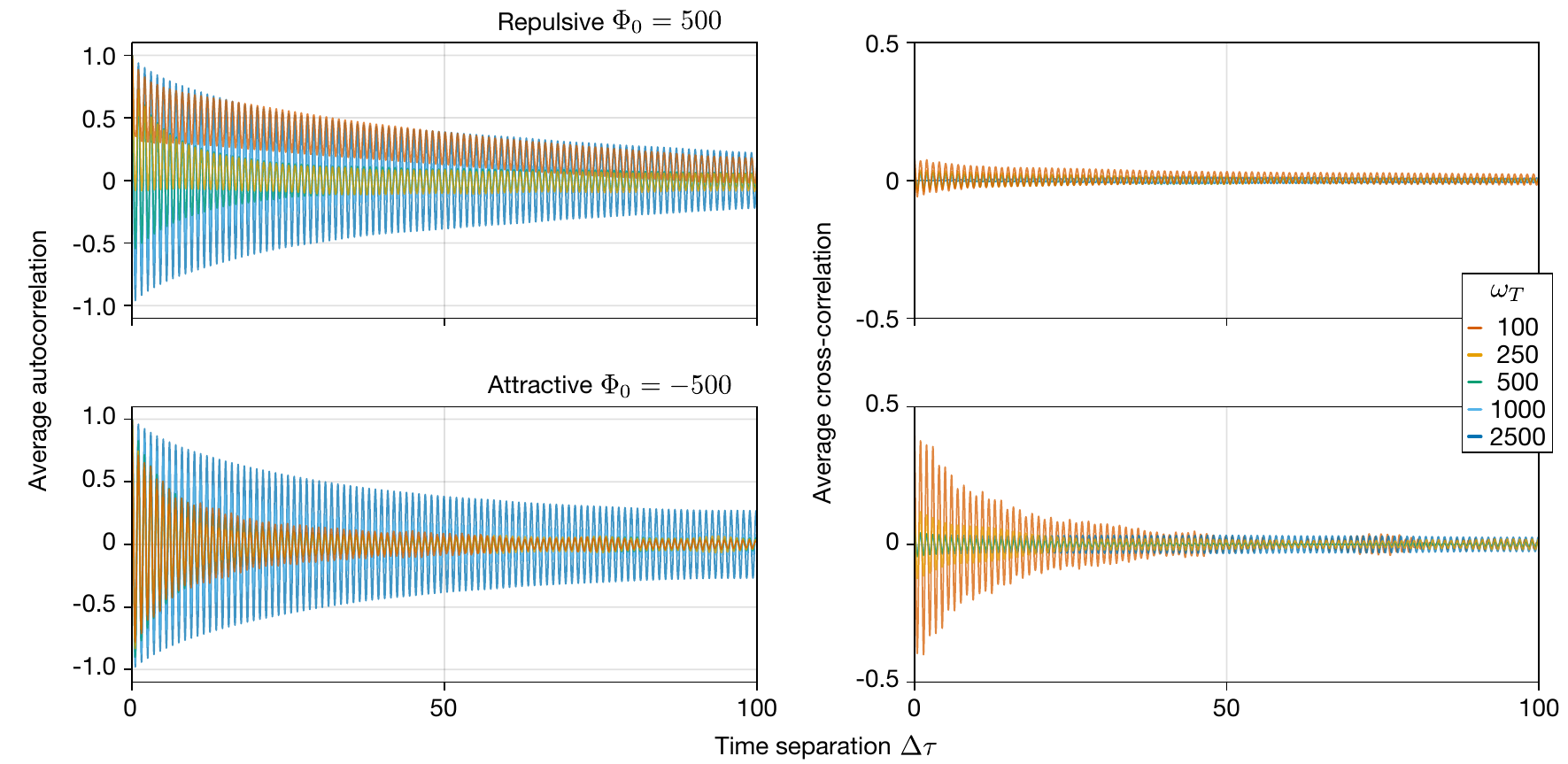}
    \caption{\textbf{Correlation functions of a multi-particle ensemble.} 
    Using the data employed to generate the left column in Fig.~\ref{fig:Energy_Distribution}, we calculate correlations between all the particle trajectories for every run.
    The auto-correlation plots are obtained by averaging 25 auto-correlations for each run, while the cross-correlations are the averages of the 600 cross-correlations.
    The auto-correlation exhibits a decaying oscillatory behavior with the decay being the consequence of the chain-induced perturbation.
    The auto-correlation for repulsive interaction at low temperatures does not change sign because the mobile particle never passes to the other side of the chain mass.
    }
    \label{fig:Correlation}
\end{figure*}

\section{Correlations}
\label{sec:Correlation}

To assess the validity of treating the mobile particles in the ensemble as independent, we calculate the position correlations for all the runs in the left column of Fig.~\ref{fig:Energy_Distribution}.
We then average the 25 autocorrelation and 600 cross-correlation functions for each $\omega_T$ and $\Phi_0$ and plot the results in Fig.~\ref{fig:Correlation}.
We observe that the auto-correlation functions exhibit an expected decaying behavior, with the decay rate generally being slower for higher temperatures.
This decay rate dependence on temperature can be attributed to a larger speed with which the mobile particles pass the chain mass, leading to a smaller perturbation.
The cross-correlations are much smaller than the auto-correlations.
The only exception is the lowest temperature for the attractive interaction, where the mobile particles fall into the Gaussian well and oscillate together.
The small magnitude of the cross-correlations in the relevant simulation run warrants the treatment of the mobile particles as independent.

%

\end{document}